\documentclass[orivec,a4paper]{llncs}
\usepackage{xspace}
\usepackage{amsmath}
\usepackage{amsfonts}
\usepackage{amssymb}
\usepackage{stmaryrd}
\SetSymbolFont{stmry}{bold}{U}{stmry}{m}{n}
\usepackage{times}
\usepackage{mathrsfs}
\usepackage{multirow}
\usepackage{cite}
\usepackage{booktabs}
\usepackage{wrapfig}
\usepackage{paralist}
\usepackage{tikz}
\usetikzlibrary{automata,arrows,intersections,positioning,calc,shapes}
\usepackage{bigstrut}
\usepackage{dcolumn}
\newcolumntype{d}{D{,}{,}{3.3}}

\newcommand{\prism}{\textsc{PRISM}\xspace}
\newcommand{\jltltwodstar}{\textsc{(J)Ltl2Dstar}\xspace}
\newcommand{\mrmc}{\textsc{MRMC}\xspace}
\newcommand{\iscasmc}{\textsc{IscasMC}\xspace}
\newcommand{\liquor}{\textsc{LiQuor}\xspace}

\newcommand{\buchi}{B\"{u}chi\xspace}

\newcommand{\mc}{\mathcal{M}}
\newcommand{\aut}{\mathcal{A}}
\newcommand{\graph}{\mathcal{G}}

\newcommand{\bigO}{\mathcal{O}}

\newcommand{\cyl}{\mathit{Cyl}}
\newcommand{\first}{\mathit{fst}}

\DeclareMathOperator{\proj}{Proj}

\newcommand{\states}{S}
\newcommand{\init}{\bar{s}}
\newcommand{\labelFunc}{L}
\newcommand{\pmat}{\mathbf{P}}
\newcommand{\lpmat}{\pmat_l}
\newcommand{\upmat}{\pmat_u}

\newcommand{\gmat}{\mathbf{\Delta}}

\newcommand{\ACC}{\mathit{ACC}}
\newcommand{\tranFunct}{\mathbf{T}}

\newcommand{\lang}{\mathcal{L}}

\newcommand{\prob}{\mathfrak{P}}

\newcommand{\varSet}{V}
\newcommand{\rationalFunctions}[1][\varSet]{\mathcal{F}_{#1}}
\newcommand{\eval}{\upsilon}
\DeclareMathOperator{\dom}{Dom}

\newcommand{\logicTrue}{\mathit{true}}

\newcommand{\logicUntil}{\mathsf{U}}
\newcommand{\logicNext}{\mathsf{X}}
\newcommand{\logicFinally}{\mathsf{F}}
\newcommand{\logicGlobally}{\mathsf{G}}
\newcommand{\logicProb}{\mathbb{P}}

\newcommand{\ltlGF}{\logicGlobally\logicFinally}
\newcommand{\ltlFG}{\logicFinally\logicGlobally}
\newcommand{\logicEventually}{\lozenge}
\newcommand{\logicAlways}{\square}

\newcommand{\satisfies}{\Vdash}

\renewcommand{\phi}{\varphi}
\renewcommand{\epsilon}{\varepsilon}

\newcommand{\paths}{\mathit{Paths}}
\newcommand{\pathsfin}{\paths_\mathit{fin}}
\newcommand{\distr}{\mu}
\newcommand{\dirac}{\delta}
\newcommand{\mpath}{\pi}
\newcommand{\mpaths}{\Pi}
\newcommand{\AP}{\mathit{AP}}

\newcommand{\interval}[1]{\mathtt{#1}}

\newcommand{\intJ}{\interval{J}}

\newcommand{\natIntK}[1][k]{[1..#1]}

\newcommand{\defeq}{\overset{\mathrm{def}}{=}}
\newcommand{\el}{\mathit{el}}

\newcommand{\reals}{\mathbb{R}}
\newcommand{\posreals}{\reals^{\geq 0}}

\newcommand{\naturals}{\mathbb{N}}

\newcommand{\setcond}[2]{\{\, #1 \mid #2 \,\}}
\newcommand{\setnocond}[1]{\{#1\}}
\newcommand{\family}[2]{{\setnocond{#1}}_{#2}}

\newcommand{\Disc}[1]{\mathrm{Disc}(#1)}
\newcommand{\SubDisc}[1]{\mathrm{SubDisc}(#1)}
\newcommand{\Supp}[1]{\mathrm{Supp}(#1)}

\newcommand{\sigmafield}{\mathcal{F}}

\newcommand{\insym}{\mathop{\mathrm{Inf}}}
\newcommand{\posscc}{\mathop{\mathrm{pos}}}
\newcommand{\negscc}{\mathop{\mathrm{neg}}}

\newcommand{\head}{\mathit{head}}
\newcommand{\tail}{\mathit{tail}}

\begin{document}
\title{An Efficient Synthesis Algorithm for Parametric Markov Chains Against Linear Time Properties%
\thanks{We have very recently noticed the paper accepted at CAV~\cite{BaierKKKMW16} which treats the non-parametric Markov chains using a similar approach. Our approach has been developed in parallel.}}

\author{Yong Li\inst{1,2} \and Wanwei Liu\inst{3}
	\and Andrea Turrini\inst{1} \and Ernst Moritz Hahn\inst{1} \and Lijun Zhang\inst{1}}
\institute{State Key Laboratory of Computer Science,
	Institute of Software, CAS, China
\and University of Chinese Academy of Sciences, China
\and College of Computer Science,
	National University of Defense Technology,
	China}

\maketitle
\begin{abstract}
In this paper, we propose an efficient algorithm for the parameter synthesis of PLTL formulas with respect to parametric Markov chains.
The PLTL formula is translated to an almost fully partitioned B\"uchi automaton which is then composed with the parametric Markov chain.
We then reduce the problem to solving an optimisation problem, allowing to decide the satisfaction of the formula using an SMT solver.
The algorithm works also for interval Markov chains.
The  complexity is linear in the size of the Markov chain, and exponential in the size of the formula.
We provide a prototype and show the efficiency of our approach on a number of benchmarks.
\end{abstract}

\section{Introduction}
\label{sec:intro}
Model checking, an automatic verification technique, has attracted much attention as it can be used to verify the correctness of software and hardware systems~\cite{0007403,Clarke08,0020348}.
In classical model checking, temporal formulas are often used to express properties that one wants to check.

Probabilistic verification problems have been studied extensively in recent years.
Markov chains (MCs) are a prominent probabilistic model used for modelling probabilistic systems.
Properties are specified using probabilistic extensions of temporal logics such as probabilistic CTL (PCTL)~\cite{HaJo94} and probabilistic LTL (PLTL)~\cite{BiancoA95} and their combination PCTL*.
In the probabilistic setting, most of the observations about CTL and LTL carry over to their probabilistic counterpart.
An exception is the complexity for verifying PLTL:
here one could have a double exponential blowup.
This is the case, because in general nondeterministic \buchi automata cannot be used directly to verify LTL properties, as they will cause imprecise probabilities in the product.
In turn, it is often necessary to construct their deterministic counterparts in terms of other types of automata, for instance Rabin or Parity automata, which adds another exponential blowup.
As a result, most of the work in literature focuses on branching time verification problems.
Moreover, state-of-the-art tools such as \prism~\cite{KwiatkowskaNP11} and \mrmc~\cite{KatoenZHHJ11} can handle large systems with PCTL specifications, but rather small systems --if at all-- for PLTL specifications.

In the seminal paper by Courcoubetis and Yannakakis~\cite{CY95}, it is shown that for MCs the PLTL model checking problem is in PSPACE.
They perform transformations of the Markov chain model recursively according to the LTL formula.
At each step, the algorithm replaces a subformula rooted at a temporal operator with a newly introduced proposition;
meanwhile, it refines the Markov chain with that proposition, and such refinement preserves the distribution.
Then, it is finally boiled down to the probabilistic model checking upon a propositional formula.
At the refinement step the state space is doubled, thus resulting in a PSPACE algorithm.
Even if it is theoretically a single exponential algorithm for analysing MCs with respect to PLTL, it has not been exploited in the state-of-the-art probabilistic model checkers.

In automata-based approaches, one first translates the LTL formula into a \buchi automaton and then analyses the product of the MC and the \buchi automaton.
This is sufficient for qualitative properties, i.e., to decide whether the specification is satisfied with probability $1$.
For quantitative properties, the \buchi automaton needs to be further transformed into a deterministic variant.
Such a determinisation step usually exploits Safra's determinisation construction~\cite{Safra/88/Safra}.
Several improvements have been made in recent years, see for instance~\cite{Piterman/07/Parity,Schewe/09/determinise,LiuW09}.
Model checkers such as \prism~\cite{KwiatkowskaNP11} and \liquor~\cite{DBLP:conf/qest/CiesinskiB06} handle PLTL formulas by using
off-the-shelf tools (e.g. \jltltwodstar \cite{KleinB07}) to perform this determinisation step.
To avoid the full complexity of the deterministic construction, Chatterjee et al.~\cite{ChatterjeeGK13} have proposed an improved algorithm for translating the formulas of the $\ltlFG$-fragment of LTL to an extension of Rabin automata.
Recently~\cite{Esparza14}, this algorithm has been extended to the complete LTL.
Despite the above improvements, the size of the resulting deterministic automaton is still the bottleneck of the approach for linear temporal properties.
In~\cite{CSS03}, it is first observed that the second blowup can be
circumvented by using \emph{unambiguous \buchi automata} (UBAs)
\cite{CartonM03}.  The resulting algorithm has the same complexity as
the one in~\cite{CY95}. Despite the importance of probabilistic model
checking, unfortunately, the algorithm in \cite{CSS03} is less
recognised. To the best of the authors knowledge, it is not applied in
any of the existing model checkers.
Recently, in\cite{Kini15}, the authors construct the so called \emph{limit deterministic}
\buchi automata that are exponential in the size of LTL\textbackslash{}GU formula $\phi$, which is another fragment of LTL.
The approach is only applied to the analysis of qualitative PLTL of the form $\logicProb_{>0}[\phi]$.

In this paper, we present a further improvement of the solution
proposed in \cite{CSS03}, adapted directly to solving the parameter synthesis problem for parametric
Markov chains. We exploit a simple construction translating the given
LTL formula to a reverse deterministic UBA, and then build the product of the parametric Markov chains.
We then extract an equation system from the product, then the synthesis problem
reduces to the existence of a solution of the equation system.
Further, we remark that the related interval Markov chains can be handled by our approach as well.
We integrate our approach in the model checker
\iscasmc~\cite{HahnLSTZ14}, and employ SMT solver to solving the obtained equation system.
We present detailed experimental results, and observe that our implementation can deal with some real-world probabilistic systems modelled by parametric Markov chains.

\paragraph{Related Work.}
In \cite{HahnHZ11}, they first use state elimination to compute the reachability
probability for parametric Markov models.
This has be improved by Dehnert \emph{et al.}\cite{Dehnert15}.
Another related models is interval Markov chains, which can be interpreted as a family of Markov chains \cite{Chatterjee08,Katoen12} whose transition probabilities lie within the interval ranges.
In \cite{Chatterjee08},
they also considered model checking $\omega$-regular properties with interval
Markov chains. They showed that the synthesis of interval Markov chains problem
against $\omega$-PCTL is decidable in PSPACE. In \cite{Katoen12}, they considered
interval Markov chains as abstraction models by using three-valued abstraction for
Markov chains.
To our best knowledge, it is
the first time that one can easily integrate parameter synthesis algorithm that is exponential in
the size of LTL formulas over parametric Markov chains.

\section{Preliminaries}
\label{sec:pre}

Given a set $W$, we say that an infinite sequence $\varpi = w_{0} w_{1} \dots$ is an \emph{$\omega$-word} if $\varpi \in W^{\omega}$.
Given a finite word $\nu = v_{0} \dots v_{k}$ and a finite or infinite word $\varpi = w_{0} w_{1} \dots$, we denote by $\nu \cdot \varpi$ the \emph{concatenation} of $\nu$ and $\varpi$, i.e., the finite or infinite word $\nu \cdot \varpi = v_{0} \dots v_{k} w_{0} w_{1} \dots$, respectively.
We may just write $\nu \varpi$ instead of $\nu \cdot \varpi$.
We denote by $\natIntK[n]$ the set of natural numbers $\setnocond{1, \cdots, n}$.

\paragraph*{Probability Theory.}

A \emph{measure} over a measurable space $(\Omega, \sigmafield)$ is a function $\mu \colon \sigmafield \to \posreals$ such that $\mu(\emptyset) = 0$ and, for each countable family $\family{\Omega_{i}}{i \in I}$ of pairwise disjoint elements of $\sigmafield$, $\mu(\cup_{i \in I} \Omega_{i}) = \sum_{i \in I} \mu(\Omega_{i})$.
If $\mu(\Omega) \leq 1$, then we call $\mu$ a \emph{sub-probability measure} and, if $\mu(\Omega) = 1$, then we call $\mu$ a \emph{probability measure}.
We say that $\mu$ is a \emph{discrete} measure over $\Omega$ if $\sigmafield$ is discrete.
In this case, for each $X \subseteq \Omega$, $\mu(X) = \sum_{x \in X} \mu(\setnocond{x})$ and we drop brackets whenever possible. 
For a set $\Omega$, we denote by $\Disc{\Omega}$ the set of discrete probability measures over $\Omega$, and by $\SubDisc{\Omega}$ the set of discrete sub-probability measures over $\Omega$.
We call $X \subseteq \Omega$ the \emph{support} of a measure $\mu$ if $\mu(\Omega \setminus X) = 0$; in particular, if $\mu$ is discrete, we denote by $\Supp{\mu}$ the minimum support set $\setcond{x \in \Omega}{\mu(x) > 0}$.
Moreover, we denote by $\dirac_{x}$, for $x \in \Omega$, the \emph{Dirac} measure such that for each $X \subseteq \Omega$, $\dirac_{x}(X) = 1$ if $x \in X$, $0$ otherwise.
If $\dirac_{x}$ is discrete, then it holds that for each $y \in \Omega$, $\dirac_{x}(y) = 1$ if $y = x$ and $\dirac_{x}(y) = 0$ if $y \neq x$.
In case $\Omega$ is countable, then the probability measure $\mu \colon \Omega \to [0,1]$ over the discrete measurable space $(\Omega, 2^{\Omega})$ can be obtained by imposing that $\sum_{x \in \Omega} \mu(x) = 1$; $\mu$ is also called a \emph{probability distribution}.

\paragraph*{Graph Theory.}
A \emph{directed graph} $G$ is a pair $G = (V, E)$ where $V$ is a finite non-empty set of \emph{vertices}, also called \emph{nodes}, and $E \subseteq V \times V$ is the set of \emph{edges} or \emph{arcs}.
Given an arc $e = (u, v)$, we call the vertex $u$ the \emph{head} of $e$, denoted by $\head(e)$, and the vertex $v$ the \emph{tail} of $e$, denoted by $\tail(e)$.
In the remainder of the paper we consider only directed graphs and we refer to them just as graphs.

A \emph{path} $\mpath$ is a sequence of edges $\mpath = e_{1} e_{2} \dots e_{n}$ such that for each $1 \leq i < n$, $\tail(e_{i}) = \head(e_{i+1})$;
we say that $v$ is \emph{reachable} from $u$ if there exists a path $\mpath = e_{1} \dots e_{n}$ such that $\head(e_{1}) = u$ and $\tail(e_{n}) = v$.

A \emph{strongly connected component} (SCC) is a set of vertices $C \subseteq V$ such that for each pair of vertices $u, v \in C$, $u$ is reachable from $v$ and $v$ is reachable from $u$;
we say that a graph $G = (V, E)$ is \emph{strongly connected} if $V$ is an SCC.
We say that an SCC $C$ is \emph{non-extensible} if for each SCC $C'$ of $G$, $C \subseteq C'$ implies $C' = C$.
Without loss of generality, in the remainder of this paper we consider only non-extensible SCCs.

We define the partial order $\preceq$ over the SCCs of the graph $G$ as follows:
given two SCCs $C_{1}$ and $C_{2}$, $C_{1} \preceq C_{2}$ if there exist $v_{1} \in C_{1}$ and $v_{2} \in C_{2}$ such that $v_{2}$ is reachable from $v_{1}$.
We say that an SCC $C$ is \emph{maximal} with respect to $\preceq$ if for each SCC $C'$ of $G$, $C \preceq C'$ implies $C' = C$.
We may call the maximal SCCs as \emph{bottom SCCs}, BSCC for short.

A graph can be enriched with labels as follows:
a \emph{labelled graph} $G$ is a triple $G = (V, \Sigma, E)$ where $V$ is a finite non-empty set of \emph{vertices}, $\Sigma$ is a finite set of \emph{labels}, and $E \subseteq V \times \Sigma \times V$ is the set of \emph{labelled edges}.
The notations and concepts on graphs trivially extend to labelled graphs.

\paragraph*{Generalized \buchi Automata.}

A \emph{generalized \buchi automaton} (GBA) $\aut$ is a tuple $\aut = (\Sigma, Q, \tranFunct, Q_{0}, \ACC)$ where 
$\Sigma$ is a finite \emph{alphabet}, 
$Q$ is a finite set of \emph{states}, 
$\tranFunct \colon Q \times \Sigma \to 2^Q$ is the \emph{transition function}, 
$Q_{0} \subseteq Q$ is the set of \emph{initial states}, 
and 
$\ACC = \setcond{F_{i} \subseteq \tranFunct}{i \in \natIntK}$ is the set of \emph{accepting sets}.

A \emph{run} of $\aut$ over an infinite word $w = a_{0} a_{1} \ldots \in \Sigma^{\omega}$ is an infinite sequence $\sigma = q_{0} a_{0} q_{1} a_{1} q_{2} \ldots \in (Q \cdot \Sigma)^\omega$ such that $q_{0} \in Q_{0}$ and for each $i \in \naturals$ it is $q_{i+1} \in \tranFunct(q_{i}, a_{i})$.
Similarly, a \emph{run} of $\aut$ over a finite word $w = a_{0} a_{1} \ldots a_{k} \in \Sigma^{*}$ is a finite sequence $\sigma = q_{0} a_{0} q_{1} a_{1} q_{2} \ldots a_{k} q_{k+1} \in Q \cdot (\Sigma \cdot Q)^{*}$ such that $q_{0} \in Q_{0}$ and for each $i \in \setnocond{0, \ldots, k}$ it is $q_{i+1} \in \tranFunct(q_{i}, a_{i})$.
Let $\insym(\sigma) = \setcond{(q,a,q') \in \tranFunct}{\forall i \in \naturals. \exists j \geq i. (q_{j}, a_{j}, q_{j+1}) = (q,a,q')}$ be the set of tuples $(q, a, q')$ occurring infinitely often in $\sigma$.
The run $\sigma$ is \emph{accepting} if $\insym(\sigma) \cap F_{i} \neq \emptyset$ for each $i \in \natIntK$.
The word $w$ is accepted by $\aut$ if there is an accepting run of $\aut$ over $w$; 
we denote by $\lang(\aut)$ the \emph{language} of $\aut$, i.e., the set of infinite words accepted by $\aut$.

Given the GBA $\aut = (\Sigma, Q, \tranFunct, Q_{0}, \ACC)$, for the sake of convenience, we denote by $\aut^{q}$ the GBA $\aut^{q} = (\Sigma, Q, \tranFunct, \setnocond{q}, \ACC)$ with initial state $q$ and accordingly for $U \subseteq Q$ we let $\aut^{U} \defeq (\Sigma, Q, \tranFunct, U, \ACC)$.

The graph $G = (V,\Sigma, E)$ underlying a GBA $\aut$ is the graph whose set of vertices (nodes) $V$ is the set of states $\states$ of $\aut$ and there is an edge $e \in E$ labelled with $a \in \Sigma$ from $q$ to $q'$ if $q' \in \tranFunct(q,a)$.
In this case, we say that $q$ is an \emph{$a$-predecessor} of $q'$ and $q'$ is an \emph{$a$-successor} of $q$.

Given a GBA $\aut$, we say that
\begin{itemize}
\item
	$\aut$ is \emph{deterministic}, if $|Q_{0}| = 1$ and $|\tranFunct(q,a)| = 1$ for each $q \in Q$ and $a \in \Sigma$;
\item
	$\aut$ is \emph{reverse deterministic} if each state has exactly one $a$-predecessor for each $a \in \Sigma$;
\item
	$\aut$ is \emph{unambiguous} if for each $q \in Q$, $a \in \Sigma$, and $q',q'' \in \tranFunct(q,a)$ such that $q' \neq q''$, we have $\lang(\aut^{q'}) \cap \lang(\aut^{q''}) = \emptyset$;
	and
\item
	$\aut$ is \emph{separated} if $\lang(\aut^q) \cap \lang(\aut^{q'}) = \emptyset$ for each pair of states $q, q' \in Q$, $q \neq q'$.
\end{itemize}

We say that a state $q \in Q$ is \emph{reenterable} if $q$ has some predecessor in $\aut$.
Let $Q'$ be the set of all reenterable states of $\aut$ and consider the GBA $\aut' = (\Sigma, Q', \tranFunct', Q', \ACC')$ where $\tranFunct' = \tranFunct|_{Q' \times \Sigma}$ and $\ACC' = \setcond{F'_{i} = F_{i} \cap \tranFunct'}{F_{i} \in \ACC, i \in \natIntK}$.
Then, we say that $\aut$ is \emph{almost unambiguous} (respectively \emph{almost separated}, \emph{almost reverse deterministic}) if $\aut'$ is unambiguous (respectively separated, reverse deterministic).

For an (almost) separated GBA $\aut$, if for each $\alpha \in \Sigma^\omega$ there exists some state $q$ of $\aut$ such that $\alpha \in \lang(\aut^q)$, then we say that $\aut$ is (almost) \emph{fully partitioned}.
Clearly, if an automaton is (almost) fully partitioned, then it is also (almost) separated, (almost) unambiguous and (almost) reverse deterministic.

\begin{wrapfigure}[7]{r}{40mm}
	\centering
	\vskip-7mm
	\begin{tikzpicture}[>=stealth',shorten >=1pt,auto]
	\path[use as bounding box] (-1.75,0.5) rectangle (1.75,1.8);
	
	\node (BA) at (0,0.5) {$\aut$};

	\node (q1) at ($(BA) + (-1.5,1)$) {$q_{1}$};
	\node (q2) at ($(BA) + (0,1)$) {$q_{2}$};
	\node (q3) at ($(BA) + (1.5,1)$) {$q_{3}$};
	
	\draw[->] ($(q1.north) + (0,0.3)$) to (q1.north);
	\draw[->] (q1) edge [bend right] node [below] {$\scriptstyle x,w$} (q2);
	\draw[->] (q2) edge [bend right] node [above] {$\scriptstyle y,z$} (q1);
	\draw[->] (q2) edge [bend right] node [below] {$\scriptstyle y$} (q3);
	\draw[->] (q3) edge [bend right] node [above] {$\scriptstyle y$} (q2);
	\end{tikzpicture}
	\caption{An example of generalised \buchi automaton}
	\label{fig:exampleAut}
\end{wrapfigure}
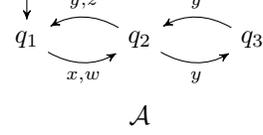
As an example of GBA that is reverse-deterministic and separated but not fully partitioned, consider the automaton $\aut$ in Fig.~\ref{fig:exampleAut}.
The fact that $\aut$ is not fully partitioned is clear since no word starting with $xw$ is accepted by any of the states $q_{1}$, $q_{2}$, or $q_{3}$.
One can easily check that $\aut$ is indeed reverse-deterministic but checking the separated property can be more involved. 
The checks involving $q_{1}$ are trivial, as it is the only state enabling a transition with label $x$ or $w$;
for the states $q_{2}$ and $q_{3}$, the separated property implies that given any $w_{1} \in \lang(\aut^{q_{2}})$, it is not possible to find some $w_{2} \in \lang(\aut^{q_{3}})$ such that $w_{1}=w_{2}$. 
For instance, suppose the number of the most front $y$'s in $w_{1}$ is odd, it must be the case that the most front $y$'s are directly followed by $x$ or $w$. 
In order to match $w_{1}$, we must choose $y$ instead of $z$ on transition $(q_{2},q_{1})$. 
It follows that the number of the most front $y$'s in $w_{2}$ is even. 
We can get similar result when the number of the most front $y$'s in $w_{1}$ is even. 
Thus $w_{1}$ and $w_{2}$ can never be the same.

\section{Parametric Markov Chains and Probabilistic LTL}
\label{sec:dtmc_mc_pre}

In this section we recall the definitions of parametric Markov chains as presented in~\cite{HahnHZ11},
interval Markov chain considered in~\cite{BenediktLW13,Chatterjee08,Katoen12} and of the logic PLTL.
In addition, we consider the translation of LTL formulas to GBAs which is used later for analysing PLTL properties.

\subsection{Parametric Markov Chains}
Before introducing the parametric Markov chain model, we briefly present some general notation.
Given a finite set $\varSet = \setnocond{x_{1}, \dotsc, x_{n}}$ with domain in $\reals$, an \emph{evaluation} $\eval$ is a partial function $\eval \colon \varSet \to \reals$.
Let $\dom(\eval)$ denote the domain of $\eval$;
we say that $\eval$ is total if $\dom(\eval) = \varSet$.
A \emph{polynomial} $p$ over $\varSet$ is a sum of monomials $p(x_{1}, \dotsc, x_{n}) = \sum_{i_{1}, \dotsc, i_{n}} a_{i_{1}, \dotsc, i_{n}} \cdot x_{1}^{i_{1}} \cdots x_{n}^{i_{n}}$ where each $i_{j} \in \naturals$ and each $a_{i_{1}, \dotsc, i_{n}} \in \reals$.
A \emph{rational function} $f$ over $\varSet$ is a fraction $f(x_{1}, \dotsc, x_{n}) = \frac{p_{1}(x_{1}, \dotsc, x_{n})}{p_{2}(x_{1}, \dotsc, x_{n})}$ of two polynomials $p_{1}$ and $p_{2}$ over $\varSet$;
we denote by $\rationalFunctions$ the set of all rational functions over $\varSet$.
Given $f \in \rationalFunctions$, $\varSet' \subseteq \varSet$, and an evaluation $\eval$, we let $f[\varSet'/\eval]$ denote the rational function obtained by replacing each occurrence of $v \in \varSet' \cap \dom(\eval)$ with $\eval(v)$.

\begin{definition}
	A \emph{parametric Markov chain} (PMC), is a tuple $\mc = (\states, \labelFunc, \init, \varSet, \pmat)$ where
	$\states$ is a finite set of \emph{states},
	$\labelFunc \colon \states \to \Sigma$ is a \emph{labelling function} where $\Sigma$ is a finite set of \emph{state labels},
	$\init \in \states$ is the \emph{initial state},
	$\varSet$ is a finite set of \emph{parameters},
	and
	$\pmat \colon \states \times \states \to \rationalFunctions$ is a \emph{transition matrix}.
\end{definition}

We now define the PMC induced with respect to a given evaluation:
\begin{definition}
	Given a PMC $\mc = (\states, \labelFunc, \init, \varSet, \pmat)$ and an evaluation $\eval$, the PMC $\mc_{\eval}$ induced by $\eval$ is the tuple $\mc_{\eval} = (\states, \labelFunc, \init, \varSet \setminus \dom(\eval), \pmat_{\eval})$ where the transition matrix $\pmat_{\eval} \colon \states \times \states \to \rationalFunctions[\varSet \setminus \dom(\eval)]$ is given by $\pmat_{\eval}(s,t) = \pmat(s,t)[\dom(\eval)/\eval]$.
\end{definition}

We say that a \emph{total} evaluation is \emph{well-defined} for a PMC $\mc$ if $\pmat_{\eval}(s,s') \in [0,1]$ and $\sum_{t \in \states} \pmat_{\eval}(s,t) = 1$ for each $s,s' \in \states$.
In the remainder of the paper we consider only well-defined evaluations and we require that, for a given PMC $\mc$ and two states $s,t \in \states$, if $\pmat_{\eval}(s,t) > 0$ for some evaluation $\eval$, then $\pmat_{\eval'}(s,t) > 0$ for the $\eval'$ considered.
We may omit the actual evaluation $\eval$ when we are not interested in the actual value for $\pmat_{\eval}(s,t)$, such as for the case $\pmat_{\eval}(s,t) > 0$.

We use $|\states|$ to denote the number of states, and $|\mc|$ for the number of non-zero probabilistic transitions, i.e., $|\mc| = |\setcond{(s,s') \in \states \times \states}{\pmat(s,s') > 0}|$.

The \emph{underlying graph} of a PMC $\mc$ is the graph $G = (V,E)$ where $V = \states$ and $E = \setcond{(s,s') \in \states \times \states}{\pmat(s, s') > 0}$.

A \emph{path} is a sequence of states $\mpath = s_{0} s_{1} \dots$ satisfying $\pmat(s_{i}, s_{i+1})>0$ for all $i\ge 0$.
We call a path $\mpath$ \emph{finite} or \emph{infinite} if the sequence $\mpath$ is finite or infinite, respectively.
We use $\mpath(i)$ to denote the suffix $s_{i} s_{i+1} \dots$
and
we denote by $\paths^\mc$ and $\pathsfin^\mc$ the set of all infinite and finite paths of $\mc$, respectively.
An infinite path $\mpath = s_{0} s_{1} \dots$ defines the $\omega$-word $w_{0} w_{1} \ldots \in \Sigma^\omega$ such that $w_{i} = \labelFunc(s_{i})$ for $i \in \naturals$.

For a finite path $s_{0} s_{1} \dots s_{k}$, we denote by $\cyl(s_{0} s_{1} \dots s_{k})$ the \emph{cylinder set} of $s_{0} s_{1} \dots s_{k}$, i.e., the set of infinite paths starting with prefix $s_{0} s_{1} \dots s_{k}$.
Given an evaluation $\eval$, we define the measure of the cylinder set by $\prob^{\mc_{\eval}} \big(\cyl(s_{0} s_{1} \dots s_{k}) \big) \defeq \dirac_{\init}(s_{0}) \cdot \prod_{i=0}^{k-1} \pmat_{\eval}(s_{i}, s_{i+1})$.
For a given PMC $\mc$ and an evaluation $\eval$, we can extend $\prob^{\mc_{\eval}}$ uniquely to a probability measure over the $\sigma$-field generated by cylinder sets~\cite{KemenySK66}.

We call the bottom SCCs of the underlying graph $G$ \emph{ergodic sets} and for each ergodic set $C$, we call each state $s \in C$ \emph{ergodic}.
A nice property of a bottom SCC $C$ is the so-called \emph{ergodicity property}:
for each $s \in C$, $s$ will be reached again in the future with probability $1$ from any state $s' \in C$, including $s$ itself.
Moreover, for each finite path $\mpath$ within $C$, $\mpath$ will be performed again in the future with probability $1$.

In this paper we are particularly interested in $\omega$-regular properties $\lang \subseteq \Sigma^\omega$ and the probability $\prob^\mc(\lang)$ for some measurable set $\lang$.
Such properties are known to be measurable in the $\sigma$-field generated by cylinders~\cite{Vardi85}.
We write $\prob_{s}^\mc$ to denote the probability function when assuming that $s$ is the initial state of the PMC $\mc$.
To simplify the notation, we omit the superscript $\mc$ whenever $\mc$ is clear from the context and we use $\mpaths$ as a synonym for $\paths$.

\subsection{Interval Markov chain}
In this section we recall the definition of interval Markov chain~\cite{Chatterjee08,Katoen12} and show
how it can be converted to a parametric Markov chain.
\begin{definition}
An \emph{interval Markov chain} (IMC) is a tuple $\mc=(\states, \labelFunc, \init, \lpmat, \upmat)$ where $\states$, $\labelFunc$ and $\init$ are as for PMCs while $\lpmat, \upmat \colon \states \times \states \to [0,1]$ are the transition matrices such that for each $s, s' \in \states$, $\lpmat(s,s') \leq \upmat(s,s')$.
\end{definition}

We show how to convert an IMC to a PMC in the following.
Given an IMC $\mc=(\states, \labelFunc, \init, \lpmat, \upmat)$, we define the corresponding PMC $\mc'=(\states, \labelFunc, \init, \pmat)$ as follows.
For every pair of states, say $(s,t)$, we add a new parameter $p_{st}$ to $\varSet$ such that
$\varSet = \setcond{p_{st}}{\lpmat(s,t) \leq p_{st} \leq \upmat(s,t)}$;
then, we define $\pmat$ as $\pmat(s,t) = p_{st}$.
For instance, suppose in an IMC, there is a state $s$ with two successors, namely $t$ and $w$, with $\lpmat(s,t) = 0.2$, $\lpmat(s, w) = 0.3$, $\upmat(s, t) = 0.7$ and $\upmat(s, w) = 0.5$.
We add two parameters $p_{st}$ and $p_{sw}$ for the pairs $(s,t)$ and $(s,w)$ whose ranges are $[0.2, 0.7]$ and $[0.3, 0.5]$ respectively.
Moreover, in order to get an instance of Markov chain from the resulting PMC, we must make sure that $p_{st} + p_{sw} = 1$.

\subsection{Probabilistic Linear Time Temporal}
\label{sec:pctl}
Throughout the whole paper, we will assume that the state space $S$ of any PMC is always equipped with labels that identify distinguishing state properties.
For this, we let $\AP$ denote a set of atomic propositions.
We assume $\Sigma = 2^\AP$ as state labels, so that $\labelFunc(s)$ specifies the subset of atomic propositions holding in state $s$.

We first recall the linear time temporal logic (LTL).
The syntax of LTL is given by:
\[
\phi \defeq p \mid \neg \phi \mid \phi \wedge \phi \mid \logicNext \phi \mid \phi \logicUntil \phi
\]
where $p \in \AP$.
We use standard derived operators, such as:
$\phi_{1} \vee \phi_{2} \defeq \neg (\neg \phi_{1} \wedge \neg \phi_{2})$, $\logicTrue \defeq a \vee \neg a$,
$\phi_{1} \rightarrow \phi_{2} \defeq \neg \phi_{1} \vee \phi_{2}$,
$\logicEventually \phi \defeq \logicTrue \logicUntil \phi$, and $\logicAlways \phi \defeq \neg( \logicEventually \neg \phi)$.
Semantics is standard and is omitted here.

A probabilistic LTL (PLTL) formula has the form $\logicProb_{\intJ}(\phi)$ where $\intJ \subseteq [0,1]$ is a non-empty interval with rational bounds and $\phi$ is an LTL formula.
In a PMC $\mc$ with evaluation $v$, for a state $s \in \states$ and a formula $\logicProb_{\intJ}(\phi)$, we have:
\begin{align}
s \models \logicProb_{\intJ}(\phi) \text{ if and only if } \prob^{\mc_{v}}_{s}(\setcond{\mpath \in \mpaths}{\mpath \models \phi}) \in \intJ
\label{eq:prob_{s}emantics}
\end{align}
where $\prob^{\mc_{v}}_{s}(\setcond{\mpath \in \mpaths}{\mpath \models \phi})$, or $\prob^{\mc_{v}}_{s}(\phi)$ for short, denotes the probability measure of the set of all paths which satisfy $\phi$.
The synthesis problem of this paper is thus to find such a $v$ if possible or to prove that the LTL formula is invalid for all valid $v$.
From the measurability of $\omega$-regular properties, we can easily show that for any PLTL path formula $\phi$, the set $\setcond{\mpath \in \mpaths}{\mpath \models \phi}$ is measurable in the $\sigma$-field generated by the cylinder sets.

\subsection{From LTL to \buchi automaton}
\label{subsubsec}
The following section describes how we can transform a given LTL formula into a GBA which has the required properties for
the subsequent model checking procedure.
\begin{definition}
\label{def:ltltoaut}
The set of \emph{elementary formulas} $\el(\phi)$ for a given LTL formula $\phi$ is defined recursively as follows:
 	$\el(p) = \emptyset$ if $p \in \AP$;
 	$\el(\neg \psi) = \el(\psi)$;
 	$\el(\phi_{1} \wedge \phi_{2}) = \el(\phi_{1}) \cup \el(\phi_{2})$;
 	$\el(\logicNext \psi) = \setnocond{\logicNext \psi} \cup \el(\psi)$;
 	and
 	$\el(\phi_{1} \logicUntil \phi_{2})= \setnocond{\logicNext(\phi_{1} \logicUntil \phi_{2})} \cup \el(\phi_{1}) \cup \el(\phi_{2})$.

Given a set $V \subseteq \el(\phi)$ and $a \in \Sigma = 2^{\AP}$, we inductively define the \emph{satisfaction relation} $\satisfies$ for each subformula of $\phi$ as follows:
\begin{alignat*}{2}
	(V,a) & {} \satisfies p &~& \text{if $p \in a$ in the case of $p \in \AP$,}\\
	(V,a) & {} \satisfies \neg \psi &~& \text{if it is not the case that $(V,a) \satisfies \psi$,}\\
	(V,a) & {} \satisfies \phi_{1} \wedge \phi_{2} &~& \text{if $(V,a) \satisfies \phi_{1}$ and $(V,a) \satisfies \phi_{2}$,}\\
	(V,a) & {} \satisfies \logicNext \psi &~& \text{if $\logicNext \psi \in V$, and}\\
	(V,a) & {} \satisfies \phi_{1} \logicUntil \phi_{2} &~& \text{if $(V,a) \satisfies \phi_{2}$ or, $(V,a) \satisfies \phi_{1}$ and $(V,a) \satisfies \logicNext(\phi_{1} \logicUntil \phi_{2})$.}
\end{alignat*}

Finally, $\aut_{\phi} = (\Sigma = 2^{\AP}, Q_{\phi}, \tranFunct_{\phi}, \setnocond{\phi}, \ACC_{\phi})$ is the \buchi automaton  where:
\begin{itemize}
\item
	$Q_{\phi} = \setnocond{\phi} \cup 2^{\el(\phi)}$;
\item
	$\tranFunct_{\phi}(\phi, a) = \setcond{V \subseteq \el(\phi)}{(V,a) \satisfies \phi}$ and for each $V \subseteq \el(\phi)$, we have:
	$\tranFunct_{\phi}(V,a) = \setcond{U \subseteq \el(\phi)}{\forall \logicNext \psi \in \el(\phi).\,\logicNext \psi \in V \Longleftrightarrow (U,a) \satisfies \psi}$;
	and
\item
	$\ACC_{\phi} = \setnocond{F_{\psi}}$ where for each subformula $\psi = \phi_{1}\logicUntil\phi_{2}$ of $\phi$, $F_{\psi} = \setcond{(U,a,V) \in \tranFunct_{\phi}}{\text{$(V,a) \satisfies \phi_{2}$ or $(V,a) \satisfies \neg\psi$}}$.
\end{itemize}
\end{definition}
In Definition~\ref{def:ltltoaut}, each formula in $\el(\phi)$ is guaranteed to be of the form $\logicNext \phi'$;
the size of $\el(\phi)$ is precisely the number of temporal operators (i.e., $\logicNext$ and $\logicUntil$) occurring in $\phi$.

\begin{theorem}[cf.~\cite{CGH94}]
\label{thm:automata_cons}
For the automaton $\aut_{\phi}$, the following holds:
\begin{enumerate}
\item
	For each infinite word $\pi \in \Sigma^\omega$, we have $\pi \models \phi$ if and only if $\pi \in \lang(\aut_{\phi})$.
\item
	More generally, for each $U \subseteq \el(\phi)$ and $\logicNext \psi \in \el(\phi)$ we have:
	$\pi \models \psi$ if and only if $\logicNext \psi \in U$, for every $\pi \in \lang(\aut^U_{\phi})$.
\end{enumerate}
\end{theorem}
It follows directly that
\begin{corollary}
\label{cor:automata_sep}
For each $U,V\subseteq \el(\phi)$, if $U\neq V$ then $\lang(\aut^U_{\phi})\cap\lang(\aut^V_{\phi})=\emptyset$.
Moreover, $\aut_{\phi}$ is both almost unambiguous and almost separated.
\end{corollary}
We observe that for each subset $U \subseteq \el(\phi)$ and each $a \in \Sigma$, there is exactly one $a$-predecessor of $U$, namely the set $\setcond{\logicNext \psi \in \el(\phi)}{(U,a) \satisfies \psi}$.
Hence, we also have the following conclusion.
\begin{corollary}
\label{cor:automata_rd}
The automaton $\aut_{\phi}$ is almost reverse deterministic and fully partitioned.
\end{corollary}
Intuitively, for any $w\in\Sigma^\omega$, we can find a state $U=\setcond{\logicNext\psi \in \el(\phi)}{w\models\psi}$ and we observe that $w\in\lang(\aut_{\phi}^U)$ by Theorem~\ref{thm:automata_cons}.
Since $\aut_{\phi}$ is already almost separated, it follows that it is also almost fully partitioned.
Note that because of the non-reenterable initial state, the automaton may not be fully partitioned, but is almost fully partitioned.

\section{Parameter Synthesis Algorithm}
\label{sec:dtmc_mc}

We consider a parametric Markov chain $\mc$ and an almost fully partitioned automaton $\aut = (\Sigma, Q, \tranFunct, Q_{0}, \ACC)$ obtained from the LTL specification, where $\lang(\aut^{q_{1}}) \cap \lang(\aut^{q_{2}}) = \emptyset$ if $q_{1}, q_{2} \in Q_{0}$ and $q_{1} \neq q_{2}$.
To simplify the notation, in the following we assume that for a given PMC $\mc$ we have $\states = \Sigma$ and $\labelFunc(s) = s$ for each $s \in \states$;
this modification does not change the complexity of probabilistic model checking~\cite{CY95}.
Below we define the product graph:
\begin{definition}
Given the automaton $\aut = (\Sigma, Q, \tranFunct, Q_{0}, \ACC)$ and the PMC $\mc = (\states, \labelFunc, \init, \varSet, \pmat)$, the \emph{product graph} of $\aut$ and $\mc$, denoted $\graph = \aut \times \mc$, is the graph $\graph = (\Gamma, \gmat)$ where $\Gamma = \setcond{(q,s)}{q \in Q, s \in \states}$ and $((q,s), (q',s')) \in \gmat$ (also written $(q,s) \gmat (q',s')$) if and only if $\pmat(s,s') > 0$ and $q' \in \tranFunct(q, \labelFunc(s))$.

Suppose that $\ACC = \setnocond{F_{1}, \dots, F_{k}}$.
We say that an SCC $C$ of $\graph$ is \emph{accepting} if for each $F_{i} \in \ACC$, there exist $(q, s), (q', s') \in C$ such that $(q,s) \gmat (q',s')$ and $(q, a, q') \in F_{i}$ for some $a \in \Sigma$.
\end{definition}

Given an SCC $C$ of $\graph$, we denote by $\mathscr{H}(C)$ the \emph{corresponding SCC} of $\mc$, where $\mathscr{H}(C) = \setcond{ s \in \states}{(q, s) \in C}$. 
We denote by $\proj$ a function to get the corresponding path of $\mc$ from the path of $\graph$, i.e., $\proj((q_{0}, s_{0})(q_{1},s_{1})\cdots) = s_{0} s_{1}\cdots$ and we usually call the path $s_{0} s_{1}\cdots$ the projection of $(q_{0}, s_{0})(q_{1},s_{1})\cdots$.
For convenience, we also write $\alpha \trianglelefteq \beta$ if the (finite) path $\alpha$ is a fragment of the (finite) path $\beta$.
\begin{definition}[Complete SCC]
For an SCC $C$ of $\graph$ and $K=\mathscr{H}(C)$ the corresponding SCC of $\mc$, we say that $C$ is \emph{complete} if for each finite path $\sigma_{K}$ in $K$,
we can find a finite path $\sigma_{C}$ in $C$ such that $\sigma_{K} = \proj(\sigma_{C})$.
\end{definition}
\begin{figure}[t]
	\centering
	\begin{tikzpicture}[>=stealth',shorten >=1pt,auto]
	
	\path[use as bounding box] (-0.5,-0.1) rectangle (11.1,3.6);
	
	\node (BA) at (0.25,0) {$\aut$};
	\node (MC) at (3.5,0) {$\mc$};
	\node (BAxMC) at (8,0) {$\aut \times \mc$};

	\node (q1) at ($(BA) + (0,3)$) {$q_{1}$};
	\node (q2) at ($(BA) + (0,2)$) {$q_{2}$};
	\node (q3) at ($(BA) + (0,1)$) {$q_{3}$};
	
	\draw[->] ($(q1.north) + (0,0.3)$) to (q1.north);
	\draw[->] (q1) edge [bend right] node [left] {$\scriptstyle x,w$} (q2);
	\draw[->] (q2) edge [bend right] node [right] {$\scriptstyle y,z$} (q1);
	\draw[->] (q2) edge [bend right] node [left] {$\scriptstyle y$} (q3);
	\draw[->] (q3) edge [bend right] node [right] {$\scriptstyle y$} (q2);

	\node (y) at ($(MC) + (-1,1)$) {$y$};
	\node (x) at ($(MC) + (-1,3)$) {$x$};
	\node (z) at ($(x) + (1.5, 0)$) {$z$};
    \node (w) at ($(y) + (1.5, 0)$) {$w$};

	\draw[->] ($(x.north) + (0,0.3)$) to (x.north);
	\draw[->] (x) edge [bend right] node [left] {$\scriptstyle $} (y);
	\draw[->] (y) edge [bend right] node [right] {$\scriptstyle $} (x);
	\draw[->] (y) edge [loop left] node [left] {$\scriptstyle $} (y);

    	\draw[->] (x) edge [] node [above] {$\scriptstyle $} (z);
    	\draw[->] (z) edge [bend left] node [right] {$\scriptstyle $} (w);
	\draw[->] (w) edge [bend left] node [right] {$\scriptstyle $} (z);
	\draw[->] (z) edge [loop right] node [above] {$\scriptstyle $} (z);

    \node (q2x) at ($(BAxMC) + (-2,1)$) {$(q_{2}, x)$};
	\node (q1y) at ($(BAxMC) + (-2,2)$) {$(q_{1}, y)$};
	\node (q3x) at ($(BAxMC) + (-2,3)$) {$(q_{3}, x)$};
	
	\node (q3y) at ($(BAxMC) + (0,1)$) {$(q_{3}, y)$};
    \node (q2y) at ($(BAxMC) + (0,2)$) {$(q_{2}, y)$};	
	\node (q1x) at ($(BAxMC) + (0,3)$) {$(q_{1}, x)$};
	\node (q1w) at ($(q3y) + (2,0)$) {$(q_{1}, w)$};
	\node (q2z) at ($(q2y) + (2,0)$) {$(q_{2}, z)$};
	\node (q1z) at ($(q1x) + (2,0)$) {$(q_{1}, z)$};

    \draw[->] ($(q1x.north) + (0,0.3)$) to (q1x.north);
    \draw[->] (q1x) edge [bend left] node [right] {} (q2y);
    \draw[->] (q1x) edge [] node [right] {} (q2z);
    \draw[->] (q2y) edge [bend left] node [left] {} (q1x);
    \draw[->] (q2y) edge [bend left] node [right] {} (q3y);
    \draw[->] (q2y) edge [] node [right] {} (q3x);
    \draw[->] (q2y) edge [] node [right] {} (q1y);
    \draw[->] (q3y) edge [] node [right] {} (q2x);
    \draw[->] (q3y) edge [bend left] node [right] {} (q2y);

    \draw[->] (q2z) edge [] node [right] {} (q1z);
    \draw[->] (q2z) edge [bend left] node [right] {} (q1w);
    \draw[->] (q1w) edge [bend left] node [left] {$\scriptstyle $} (q2z);

    \draw[dotted] ($(q1x) + (-0.55, 0.25)$) rectangle ($(q3y) + (0.55, -0.25)$);
    \node (c1) at ($(q3y) + (-0.8, 0.4)$) {$C_{1}$};

    \draw[dotted] ($(q2z) + (-0.55, 0.25)$) rectangle ($(q1w) + (0.55, -0.25)$);
    \node (c2) at ($(q1w) + (0.95, 0.4)$) {$C_{2}$};

    \draw[dotted] ($(x) + (-0.55, 0.25)$) rectangle ($(y) + (0.55, -0.25)$);
    \node (k1) at ($(y) + (-0.85, 0.4)$) {$K_{1}$};

    \draw[dotted] ($(z) + (-0.55, 0.25)$) rectangle ($(w) + (0.55, -0.25)$);
    \node (k1) at ($(w) + (0.8, 0.4)$) {$K_{2}$};

	\end{tikzpicture}
	\caption{The GBA $\aut$ from Fig.~\ref{fig:exampleAut}, a PMC $\mc$, and their product $\aut \times \mc$}
	\label{fig:exampleOfMCandBuechiAndComplete}
\end{figure}
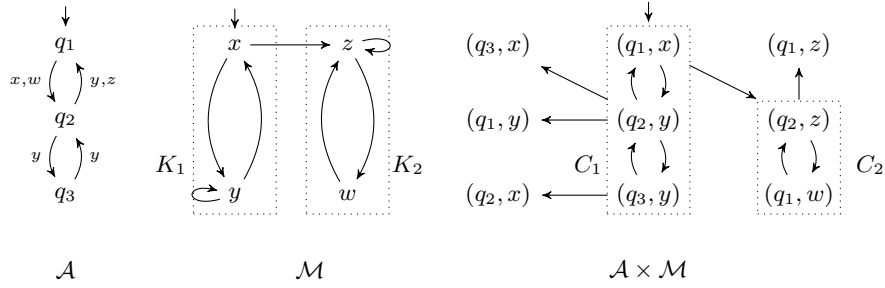
Consider the product $\aut \times \mc$ shown in Fig.~\ref{fig:exampleOfMCandBuechiAndComplete}. 
It has two non-trivial SCCs, namely $C_{1}$ and $C_{2}$. 
Clearly, $K_{1}$ and $K_{2}$ are the corresponding SCCs of $C_{1}$ and $C_{2}$, respectively. 
We observe that $C_{1}$ is a complete SCC while $C_{2}$ is not complete since the finite path $zz$ of $K_{2}$ is not a projection of any finite path in $C_{2}$.
The key observation is that some transitions in the SCCs of $\mc$ may be lost in building the
product, so we only consider the complete SCCs in the product to make sure that no transitions
of the projection SCCs are missing.

The following lemma characterises the property of a complete SCC of $\graph$.
\begin{lemma}
\label{lem:prop_completeness}
Consider a complete SCC $C$ of $\graph$ with $\mathscr{H}(C) = K$ and an arbitrary finite path $\rho_{C}$ in $C$.
Then, there exists some finite path $\sigma_{K}$ in $K$ with the following property:
for each finite path $\sigma_{C}$ in $C$ with $\sigma_{K}=\proj(\sigma_{C})$, $\sigma_{C}$ contains $\rho_{C}$ as a fragment, i.e, $\rho_{C}\trianglelefteq\sigma_{C}$.
\end{lemma}

 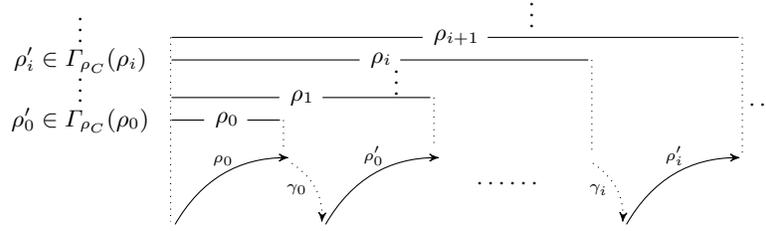
\begin{figure}
	\centering
	\begin{tikzpicture}[>=stealth',shorten >=1pt,auto]
	\path[use as bounding box] (-1.65,0.1) rectangle (8.6,3.1);
	
	\node (r0) at (0.5,0) {};
	\node (r0ga) at (2,1) {};
	\node (ga0rp) at (2.5, 0) {};
    \node (r0p) at (4, 1) {};	
	\draw[->] (r0) edge [bend left] node [above] {$\scriptstyle \rho_{0}$} ($(r0ga)+(0.1, 0)$);
    \draw[->] (r0ga) edge [dotted, bend left] node [left] {$\scriptstyle \gamma_{0}$} (ga0rp);
    \draw[->] (ga0rp) edge [bend left] node [above] {$\scriptstyle \rho'_{0}$} ($(r0p)+(0.1, 0)$);

    \node (r0l) at (0.4, 1.5) {};
    \node (r0m) at (1.25, 1.5) {$\rho_{0}$};
    \node (r0r) at (2.1, 1.5) {};
    \node (ropl) at (-0.7, 1.5) {$\rho'_{0}\in\Gamma_{\rho_{C}}(\rho_{0})$};
    \draw (r0l) -- (r0m.west);
    \draw (r0m.east) -- (r0r);

    \node (ropldots) at (-0.7, 2) {$\vdots$};

    \node (r1l) at (0.4, 1.8) {};
    \node (r1m) at (2.25, 1.8) {$\rho_{1}$};
    \node (r1r) at (4.1, 1.8) {};
    \draw (r1l) -- (r1m.west);
    \draw (r1m.east) -- (r1r);

    \node (dotspath) at (5, 0.65) {$\cdots\cdots$};

    \node (riga) at (6, 1) {};
    \node (gairp) at (6.5, 0) {};
    \node (rip) at (8, 1) {};
    \draw[->] (riga) edge [dotted, bend left] node [left] {$\scriptstyle \gamma_i$} (gairp);
    \draw[->] (gairp) edge [bend left] node [above] {$\scriptstyle \rho'_i$} ($(rip)+(0.1, 0)$);

    \node (dotsline) at (3.5, 2.1) {$\vdots$};

    \node (ril) at (0.4, 2.3) {};
    \node (rim) at (3.3, 2.3) {$\rho_i$};
    \node (rir) at (6.2, 2.3) {};
    \node (ripl) at (-0.7, 2.3) {$\rho'_i\in\Gamma_{\rho_{C}}(\rho_i)$};
    \draw (ril) -- (rim.west);
    \draw (rim.east) -- (rir);

    \node (ripldots) at (-0.7, 2.8) {$\vdots$};

    \node (ripl) at (0.4, 2.6) {};
    \node (ripm) at (4.3, 2.6) {$\rho_{i+1}$};
    \node (ripr) at (8.2, 2.6) {};
    \draw (ripl) -- (ripm.west);
    \draw (ripm.east) -- (ripr);

    \draw ($(ripl)+(0.1,0)$) edge [dotted] node {} (r0);
    \draw ($(r0r)+(-0.1,0)$) edge [dotted] node {} (r0ga);
    \draw ($(r1r)+(-0.1,0)$) edge [dotted] node {} (r0p);
    \draw ($(rir)+(-0.1,0)$) edge [dotted] node {} ($(riga)+(0.1,0)$);
    \draw ($(ripr)+(-0.1,0)$) edge [dotted] node {} ($(rip)+(0.1,0)$);

    \node (dotsupper) at (5.3, 3) {$\vdots$};  	
    \node (dotback) at (8.4, 1.7) {$\cdots$};
	\end{tikzpicture}
	\caption{Construction of $\rho_{t}$ in Lemma~\ref{lem:prop_completeness}}
	\label{fig:constructionOfRho}
\end{figure}
\begin{proof}

Clearly, $C$ must be the product of $K$ and some component of $\aut$.
Recall that $\aut$ is almost reverse-deterministic, then for each path $\alpha_{C}$ in $C$ and each finite path $\beta_{K}$ in $K$, there is at most one path $\gamma_{C} \cdot \alpha_{C}$ in $C$ with $\proj(\gamma_{C}) = \beta_{K}$.
In the following, we call $\gamma_{C} \cdot \alpha_{C}$ the \emph{$\beta_{K}$-backward-extension} of $\alpha_{C}$.

Given a finite path $\alpha$ in $C$, we define
$
	\Gamma_\alpha(\gamma) \defeq \setcond{\beta}{\proj(\beta) =\proj(\gamma), \alpha \ntrianglelefteq \beta}
$
for every finite path $\gamma$ in $C$.

Then, for the path $\rho_{C}$, we define a sequence of finite paths $\rho_{0}$, $\rho_{1}$, $\rho_{2}$, \dots{} as follows:
\begin{itemize}
\item
	$\rho_{0} = \rho_{C}$.
 According to the definition, $\rho_{0} \notin \Gamma_{\rho_{C}}(\rho_{0})$, because $\rho_{C} \trianglelefteq \rho_{C} = \rho_{0}$.
	Moreover, we have that $\Gamma_{\rho_{C}}(\rho_{0})$ is a finite set, let $\ell = |\Gamma_{\rho_{C}}(\rho_{0})| \leq |Q|^n$, where $n$ is the number of states along the finite path $\rho_{0}$ and $Q$
	is the state space of $\aut$.
\item
	For each $i \geq 0$, we impose the inductive hypothesis that $\rho_{C} \trianglelefteq \rho_{i}$.
	Then, if $|\Gamma_{\rho_{C}}(\rho_{i})| > 0$, we arbitrarily choose a path $\rho'_{i} \in \Gamma_{\rho_{C}}(\rho_{i})$.
	Since $C$ is an SCC, then there exists some path $\gamma_{i}$ connecting the last node of $\rho_{i}$ to the first state of $\rho'_{i}$.
\item
	Let $\rho_{i+1} = \rho_{i} \cdot \gamma_{i} \cdot \rho'_{i}$, then it is a finite path in $C$ and $\rho_{C} \trianglelefteq \rho_{i+1}$.
	It is immediate to see that $|\Gamma_{\rho_{C}}(\rho_{i+1})| < |\Gamma_{\rho_{C}}(\rho_{i})|$ since the set $\Gamma_{\rho_{C}}(\rho_{i+1})$ just consists of the $\proj(\rho_{i} \cdot \gamma_{i})$-backward-extensions of paths in $\Gamma_{\rho_{C}}(\rho_{i})$.
	As we have mentioned, each path has at most one such extension, and the extension for $\rho'_{i}$, namely $\rho_{i+1}$, does not belong to $\Gamma_{\rho_{C}}(\rho_{i+1})$ since it involves the fragment $\rho_{C}$.
\end{itemize}
Therefore, there must exist $t \leq \ell$ such that $|\Gamma_{\rho_{C}}(\rho_t)| = 0$.
Let $\sigma_{K} = \proj(\rho_t)$; then each finite path in $C$ with projection $\sigma_{K}$ must contain the fragment $\rho_{C}$.
\qed
\end{proof}

For example, consider the product $\aut \times \mc$ from Fig.~\ref{fig:exampleOfMCandBuechiAndComplete}. 
Given the complete SCC $C_{1}$ and $\rho_{C}=(q_{3},y)(q_{2},y)$, we find the specific $\sigma_{K}$ as follows.

\begin{enumerate}
\item 
	$\rho_{0} = \rho_{C}$. 
	First we get $\Gamma_{\rho_{C}}(\rho_{0}) = \setnocond{(q_{2},y)(q_{3},y)}$. 
	Since $|\Gamma_{\rho_{C}}(\rho_{0})| = 1 > 0$, we choose $\gamma_{0} = (q_{2},y)(q_{1},x)(q_{2},y)$ and for $\rho'_{0}$, we have no choices but $(q_{2},y)(q_{3},y)$.
\item 
	$\rho_{1} = \rho_{0} \cdot \gamma_{0} \cdot \rho'_{0}$. 
	Similarly, we need to compute $\Gamma_{\rho_{C}}(\rho_{1})$.
	Since $\proj(\rho_{1}) = yyxyy$ and one needs to visit $x$ after traversing $yy$, which is impossible to bypass $\rho_{C}$ to make it happen. 
	Then it gives us $|\Gamma_{\rho_{C}}(\rho_{1})| = 0$. 
	Therefore, we set $\sigma_{K}=yyxyy$ to be the specific finite path of $K_{1}$.
\end{enumerate}
Based on Lemma~\ref{lem:prop_completeness}, the following corollary relates the paths in the product and the projected Markov chains:
\begin{corollary}
\label{cor:mutual_inf}
Let $C$ be a complete SCC of $\graph$ and $K = \mathscr{H}(C)$; consider two infinite paths $\sigma_{C}$ in $C$ and $\sigma_{K}$ in $K$ such that $\sigma_{K}=\proj(\sigma_{C})$;
let $P_{C}$ and $P_{K}$ be the following properties:
\begin{itemize}
\item
	$P_{C}$: $\sigma_{C}$ visits each finite path in $C$ infinitely often;
\item
	$P_{K}$: $\sigma_{K}$ visits each finite path of $K$ infinitely often.
\end{itemize}
Then $P_{C}$ holds if and only if $P_{K}$ holds.
\end{corollary}
The proof of Corollary~\ref{cor:mutual_inf} can be found in the appendix.

\begin{definition}
We say that the SCC $C$ of $\graph$ is \emph{locally positive} if:
\begin{enumerate}
\item
	$C$ is accepting and complete.
\item
	$\mathscr{H}(C)$ is maximal with respect to $\preceq_{\mc}$ (it is a so-called \emph{bottom} SCC).
\end{enumerate}
\end{definition}
Consider again the example from Fig.~\ref{fig:exampleOfMCandBuechiAndComplete}. 
Assume the acceptance condition of $\aut$ is $\ACC = \setnocond{\setnocond{(q_{2}, y, q_{1})}}$; 
we observe that the SCC $C_{1}$ is both accepting and complete but not locally positive since $\mathscr{H}(C_{1}) = \setnocond{x,y}$ is not a bottom SCC in $\mc$.

According to Corollary~\ref{cor:mutual_inf}, the ergodicity property of Markov chains, and the definition of \buchi acceptance, we have the following result.
\begin{proposition}
\label{prop:prob_emptiness}
$\prob^\mc(\lang(A))\neq 0$ if and only if there exists some locally positive SCC in $\graph$.
\end{proposition}

For a given SCC, in order to decide whether it is locally positive, we have to judge whether it is complete.
In general, doing so is a nontrivial task.
Thanks to~\cite[Lemma~5.10]{CY95}, completeness can be checked efficiently:
\begin{lemma}
\label{lem:comp_check}
If $\aut$ is (almost) reverse deterministic, then the following two conditions are equivalent:
\begin{enumerate}[i)]
\item\label{completeArcs}
	$C$ is complete, i.e., each finite path of $\mathscr{H}(C)$ is a projection of some finite path in $C$.
\item\label{firstSCCs}
	There is no other SCC $C'$ of $\graph$ with $\mathscr{H}(C') = \mathscr{H}(C)$ such that $C' \preceq C$.
\end{enumerate}
\end{lemma}
Intuitively, in the product $\graph$ composed by an almost reverse deterministic automaton $\aut$ and a PMC $\mc$, the complete SCCs must be the SCCs whose preceded SCCs should not have the same projections. 
The detailed proof can be found in the appendix.


We now turn to the problem of computing the exact probability.
\begin{theorem}
\label{thm:ltlConcreteProbs}
Given a PMC $\mc$ and a fully partitioned \buchi automaton $\aut$, let $\graph = \aut \times \mc$ be their product.
Let $\posscc(\graph)$ be the set of all locally positive SCCs of $\graph$ and $\negscc(\graph)$ be the set of all BSCCs of $\graph$ which are not locally positive.
Further, for an SCC $C$ let $C_{\mc} = \setcond{s \in S}{\exists q \in Q .\ (q,s) \in C}$ denote the set of states of $\mc$ occurring in $C$.
We define the following equation system:
\begin{alignat}{2}
\label{eqn:cond_ind}
\distr(q,s) & {} = \sum_{s'\in S} \left( \pmat(s,s') \cdot \sum_{(q,s) \gmat (q',s')} \distr(q',s') \right) & \kern-15mm \text{$\forall q \in Q, s \in S$} \\
\label{eqn:cond_positive}
\sum_{\substack{q \in Q \\ (q,s) \in C}} \distr(q,s) & {} = 1 & \kern-15mm \text{$\forall C \in \posscc(\graph), s \in C_{\mc}$}\\
\label{eqn:cond_zeron}
\mu(q,s) & {} = 0 & \kern-15mm \text{ $\forall C \in \negscc(\graph)$ and $(q,s)\in C$}
\end{alignat}
Then, it holds that $\prob^{\mc_{\eval}}(\lang(\aut)) = \sum_{q_{0} \in Q_{0}} \distr(q_{0},\init)$ for any well-defined evaluation $\eval$.
\end{theorem}

In general, all locally positive SCCs can be projected to the BSCCs in the induced MC $\mc_{\eval}$.
In the original MC $\mc_{\eval}$, the reachability probability of every state in the accepting BSCC should be $1$.
Thus in a locally positive SCC of $\graph$, the probability mass $1$ distributes on the states in which they share the same second component, i.e, $s$ from state $(q, s)$.
This follows from the fact that the resulting product $\graph$ is almost fully partitioned so that the probability is also partitioned.

\begin{figure}[t]
	\centering
	\begin{tikzpicture}[->,>=stealth',shorten >=1pt,auto]
	
	\node (BA) at (0.5,0) {$\aut$};
	\node (MC) at (3.5,0) {$\mc$};
	\node (BAxMC) at (8,0) {$\aut \times \mc$};

	\node (q5) at ($(BA) + (0,1)$) {$q_{5}$};
	\node (q4) at ($(q5) + (1,1.5)$) {$q_{4}$};
	\node (q3) at ($(q5) + (-1,1.5)$) {$q_{3}$};
	\node (q2) at ($(q4) + (0,1.5)$) {$q_{2}$};
	\node (q1) at ($(q3) + (0,1.5)$) {$q_{1}$};
	
	\draw ($(q1.north) + (0,0.3)$) to (q1.north);
	\draw (q1) to node [left] {$\scriptstyle x$} (q3);
	\draw (q3) to node [left] {$\scriptstyle y$} (q5);
	\draw (q5) to node [right] {$\scriptstyle w$} (q1);
	\draw ($(q2.north) + (0,0.3)$) to (q2.north);
	\draw (q2) to node [right] {$\scriptstyle x$} (q4);
	\draw (q4) to node [right] {$\scriptstyle z$} (q5);
	\draw (q5) to node [left] {$\scriptstyle w$} (q2);
	
	\node (w) at ($(MC) + (0,1)$) {$w$};
	\node (y) at ($(w) + (-1,1.5)$) {$y$};
	\node (z) at ($(w) + (1,1.5)$) {$z$};
	\node (x) at ($(y) + (1,1.5)$) {$x$};
	\draw ($(x.north) + (0,0.3)$) to (x.north);
	\draw (x) to node [left, near end] {$\scriptscriptstyle 0.5 + \epsilon$} (y);
	\draw (x) to node [right, near end] {$\scriptscriptstyle 0.5 - \epsilon$} (z);
	\draw (y) to node [left, very near end] {$\scriptscriptstyle 1$} (w);
	\draw (z) to node [right, very near end] {$\scriptscriptstyle 1$} (w);
	\draw (w) to node [right, near end] {$\scriptscriptstyle 1$} (x);
	
	\node (q5w) at ($(BAxMC) + (0,1)$) {$(q_{5}, w)$};
	\node (q4y) at ($(q5w) + (2.25,1.5)$) {$(q_{4}, y)$};
	\node (q4z) at ($(q5w) + (1.25,1.5)$) {$(q_{4}, z)$};
	\node (q3y) at ($(q5w) + (-1.25,1.5)$) {$(q_{3}, y)$};
	\node (q3z) at ($(q5w) + (-2.25,1.5)$) {$(q_{3}, z)$};
	\node (q1x) at ($(q3y) + (0,1.5)$) {$(q_{1}, x)$};
	\node (q2x) at ($(q4z) + (0,1.5)$) {$(q_{2}, x)$};
	
	\draw (q1x) to (q3z);
	\draw (q1x) to (q3y);
	\draw (q3y) to (q5w);
	\draw (q5w) to (q1x);
	\draw (q2x) to (q4y);
	\draw (q2x) to (q4z);
	\draw (q4z) to (q5w);
	\draw (q5w) to (q2x);
	
	\end{tikzpicture}
	\caption{An example of a GBA $\aut$, a PMC $\mc$, and their product $\aut \times \mc$}
	\label{fig:exampleOfMCandBuechi}
\end{figure}

\begin{example}
Consider the PMC $\mc$ and the automaton $\aut$ as depicted in Fig.~\ref{fig:exampleOfMCandBuechi}, together with their product graph.
For clarity, we have omitted the isolated vertices like $(q_{1},w)$ and $(q_{5},y)$, i.e., the vertices with no incoming or outgoing edges.
One may check that $\aut$ is indeed separated, unambiguous, and reverse deterministic.
The product of $\mc$ and $\aut$ consists of a single locally positive SCC $C = \setnocond{(q_{1}, x), (q_{2}, x), (q_{3}, y), (q_{4}, z), (q_{5}, w)}$.
We state the relevant part of the equation system resulting from equations \eqref{eqn:cond_ind} and \eqref{eqn:cond_positive} of Theorem~\ref{thm:ltlConcreteProbs}:
\begin{center}
\begin{tikzpicture}
\path[use as bounding box] (-3.8,-2.5) rectangle (8.25,2);
\node (eqn:cond_ind) at (0,0) {
\begin{minipage}{76mm}
\centering
\begin{align*}
\distr(q_{3},z) & = 0\\
\distr(q_4,y) & = 0\\
\distr(q_{1},x) & = (0.5 + \epsilon) \cdot \distr(q_{3},y) + (0.5 - \epsilon) \cdot \distr(q_{3},z) \\
\distr(q_{2},x) & = (0.5 - \epsilon) \cdot \distr(q_4,z) + (0.5 + \epsilon) \cdot \distr(q_4,y) \\
\distr(q_{3},y) & = 1 \cdot \distr(q_5,w)\\
\distr(q_4,z) & = 1 \cdot \distr(q_5,w)\\
\distr(q_5,w) & = 1 \cdot (\distr(q_{1},x) + \distr(q_{2},x))
\end{align*}
\eqref{eqn:cond_ind}
\end{minipage}
};
\node (eqn:cond_positive) at (6.5,0) {
\begin{minipage}{34mm}
\centering
\begin{align*}
\\
\\
\distr(q_{1},x) + \distr(q_{2},x) & = 1 \\
\distr(q_{3},y) & = 1 \\
\distr(q_4,z) & = 1 \\
\distr(q_5,w) & = 1 \\
\end{align*}
\eqref{eqn:cond_positive}
\end{minipage}
};
\end{tikzpicture}
\end{center}
We remark that the values for the nodes $(q_{3},z)$ and $(q_4,y)$ as well as all isolated nodes like $(q_{1},w)$ or $(q_{5},y)$ are $0$, because for them the inner summation in \eqref{eqn:cond_ind} is over the empty set.
The family of solutions of this equation system has as non-zero values $\distr(q_{1},x) = 0.5 + \epsilon$, $\distr(q_{2},x) = 0.5 - \epsilon$, $\distr(q_{3},y) = 1$, $\distr(q_4,z) = 1$, and $\distr(q_5,w) = 1$.
From this, we have $\prob^\mc(\lang(\aut)) = \distr(q_{1},x) + \distr(q_{2},x) = 1$ for any well-defined evaluation $\eval$, i.e., an evaluation such that $\eval(\epsilon) \in (-0.5,0.5)$.
\end{example}

Let us summarise this section with the following results:
given a parametric Markov chain $\mc$ and an LTL formula $\phi$, both the emptiness checking and the quantitative-correctness computation could be done within time $\bigO(|\mc| \cdot 2^{|\el(\phi)|})$.
When the specification is given as an almost fully partitioned automaton $\aut$, the time complexity is $\bigO(|\mc| \cdot |\aut|)$.

\begin{remark}
We note that in \cite{BenediktLW13}, a single exponential algorithm is presented for model checking PLTL properties on Markov chains.
However, the approach presented there contains a flaw, which has been fixed in a subsequent report~\cite{BenediktLW14arxiv}.
In this fix, the authors have also exploited UBAs.
Comparing to our approaches, their corrected version has a higher complexity.
In particular, they reduce computing the probability of a language being accepted by an UBA to solving a system of linear equations encoding the probability of the languages accepted by a non-deterministic finite automaton.
The resulting complexity is indeed polynomial in the size of the UBA and of the Markov chain,
but it is higher than the one of the algorithm we consider in this work: for instance, it involves checking
accepting states which is cubic in the size of the system, whereas our approach is linear.
Moreover, the fix was not implemented in their tool tulip \cite{BenediktLW13}.
\end{remark}

\section{Experiment Results}
\label{sec:experiments}
We have implemented our synthesis algorithm for LTL properties of parametric Markov chains in our tool \iscasmc using an explicit state-space representation.
The machine we used for the experiments is a 3.6 GHz Intel Core i7-4790 with 16 GB 1600 MHz DDR3 RAM of which 12 GB assigned to the tool;
the timeout has been set to 30 minutes.
We considered three models, namely the Bounded Retransmission Protocol (BRP)~\cite{HelminkSV94} in the version of~\cite{DArgenioJJL01}, Randomized Protocol for Signing Contracts (RPSC)~\cite{Even85} and the Crowds protocol for anonymity~\cite{RR98}.
We have replaced the probabilistic choices by parametric choices to obtain the admissible failure probabilities.

\begin{table}[t]
	\centering
	\caption{Experimental Results for Parametric Markov chain models}
	\label{tab:paramtable}
	\setlength{\tabcolsep}{5pt}
	\resizebox{1\textwidth}{!}{%
		\begin{tabular}{ldr|rrrr|rrr|r}
		\toprule
		\multicolumn{1}{c}{Model (Constants)} & \multicolumn{1}{c}{\multirow{2}{*}{Constants}} & \multicolumn{1}{c|}{\multirow{2}{*}{$|\states_{\mc}|$}} & \multicolumn{1}{c}{\multirow{2}{*}{$|V_{\graph}|$}} & \multicolumn{1}{c}{\multirow{2}{*}{$\mathit{SCC}_{\graph}$}} & \multicolumn{1}{c}{\multirow{2}{*}{$\mathit{SCC}_{\mathit{pos}}$}} & \multicolumn{1}{c|}{\multirow{2}{*}{$T_{\graph}$}} & \multirow{2}{*}{$|\mathit{Vars}_{Z3}|$} & \multirow{2}{*}{$|\mathit{Cons}_{Z3}|$} & \multicolumn{1}{c|}{\multirow{2}{*}{$T_{\mathit{Z3}}$}} & \multirow{2}{*}{$T_{\mathit{mc}}$} \\ 
		\multicolumn{1}{c}{Property} & & & & & & & & & \\ 
		\midrule
		\multirow{8}{*}{
		\begin{tabular}{l}
		Crowds (TotalRuns, CrowdSize)\\
		\\
		$\logicProb\geq 0.9[ \ltlGF(\mathit{newInstance} \land $\\
		$\mathit{runCount}=0 \land\mathit{observe0} \ge 1)] $
		\end{tabular}
		}
		& 2,50 & 20534 & 20535 & 1372 & 1 & $<$1  & 314 & 320  & 13 & 13 \\
		& 2,60 & 28446 & 28447 & 1938 & 1 & $<$1  & 374 & 380  & 42 & 43\\
		& 2,70 & 39131 & 39132 & 2613 & 1 & $<$1  & 434 & 440 & 80& 82 \\
		& 2,80 & 51516 & 51517 & 3388 & 1 & 1  & 494 & 500  & 92 & 94 \\ 
		& 2,90 & 65601 & 65602 & 4263 & 1 &  1   & 554 & 560& 162 & 164 \\
		& 2,100 & 81386 & 81387 & 5238 & 1 & 2   & 614 & 620 & 198& 202  \\
		& 2,110 & 98871 & 98872 & 6313 & 1 &  2  & 674 & 680& 404 & 407 \\
		& 2,120 & 118056 & 118057 & 7488 & 1 & 2 & 734 & 740 & 436  & 439 \\
		\midrule
		\multirow{11}{*}{
		\begin{tabular}{l}
		BRP (N, MAX)\\
		\\
		$\logicProb\geq 0.9[\ltlGF(\mathit{s}=5\land\mathit{T})]$
		\end{tabular}
		}
		& 512,80 & 1084476 & 1085505 & 1111 & 0  & 15 & NE & NE& NE & 16\\
		& 512,100 & 1351476 & 1352505 & 1131 & 0  & 22 & NE & NE& NE & 22\\
		& 512,120 & 1618476 & 1619505 & 1151 & 0  & 41 & NE & NE& NE & 42 \\
		& 512,140 & 1885476 & 1886505 & 1171 & 0  & 54 & NE & NE& NE & 54 \\
		& 512,160 & 2152476 & 2153505 & 1191 & 0  & 75 & NE & NE& NE & 76 \\
		& 512,180 & 2419476 & 2420505 & 1211 & 0  & 87 & NE & NE& NE & 87 \\
		& 512,200 & 2686476 & 2687505 & 1231 & 0  & 110 & NE & NE & NE& 111 \\
		& 512,220 & 2953476 & 2954505 & 1251 & 0  & 136 & NE & NE & NE& 137 \\
		\midrule
		\multirow{8}{*}{
		\begin{tabular}{l}
		RPSC (N, L) \\
		\\
		$\logicProb\geq0.9[\ltlGF( \neg \mathit{knowA} \land \mathit{knowB})]$
		\end{tabular}
		}
		& 5,15  & 300030 & 305492 & 1 & 0& 5 & NE & NE & NE & 5 \\
		& 5,20 & 402430 & 409442 & 1  & 0& 9 & NE & NE& NE & 9 \\
		& 5,25  & 504830 & 513392 & 1 & 0& 12  & NE & NE& NE & 12 \\
		& 5,30 & 607230 & 617342 & 1  & 0& 19 & NE & NE& NE & 19  \\
		& 5,35  & 709630 & 721292 & 1 & 0& 25  & NE & NE& NE  & 25 \\
		& 5,40 & 812030 & 825242 & 1  & 0& 32 & NE & NE& NE & 32 \\
		& 5,45 & 914430 & 929192 & 1  & 0& 39  & NE & NE& NE & 39 \\
		& 5,50 & 1016830 & 1033142 & 1& 0& 46  & NE & NE& NE & 47 \\

		\bottomrule
		\end{tabular}
	}
\end{table}
In our implementation, we use Z3~\cite{DeMoura08} to solve the equation system of Theorem~\ref{thm:ltlConcreteProbs} because by using parametric transition probabilities the equation system from Theorem~\ref{thm:ltlConcreteProbs} becomes nonlinear.
Performance results of the experiments are shown in Tables~\ref{tab:paramtable} to~\ref{tab:imctable}, where the columns have the following meaning:
In column ``Model'', the information about the model, the name of the constants influencing the model size, and the analysed property is provided;
``Constants'' contains the values for the constants defining the model;
``$|\states_{\mc}|$'' and``$|V_{\graph}|$'' denote the number of states and vertices in $\mc$ and $\graph$, respectively;
``$\mathit{SCC}_{\graph}$'' reports the number of non-trivial SCCs checked in $\graph$ out of which ``$\mathit{SCC}_{\mathit{pos}}$'' are the positive ones;
and ``$T_{\graph}$'' is the time spent by constructing and checking the product graph;
``$|\mathit{Vars}_{Z3}|$'', ``$|\mathit{Cons}_{Z3}|$'' and ``$T_{\mathit{Z3}}$'' record the number of variables and constraints of the equation system we input into Z3 and its solution time;
and 
``$T_{\mathit{mc}}$'' gives the total time spent for constructing and analysing the product $\graph$ and solving the equation system.
In the tables, entries marked by ``NE'' mean that Z3 has not been executed as there were no locally positive SCCs, thus the construction and evaluation of the equation system can be avoided;
entries marked by ``--'' mean that the operation has not been performed since the analysis has been interrupted in a previous stage;
the marks ``TO'' and ``MO'' stand for a termination by timeout or memory out.
\begin{table}[t]
	\centering
	\caption{Experimental Results for Parametric Markov chain models}
	\label{tab:paramtable2}
	\setlength{\tabcolsep}{3pt}
	\resizebox{1\textwidth}{!}{%
		\begin{tabular}{ldr|rrrr|rrr|r}
		\toprule
		\multicolumn{1}{c}{Model (Constants)} & \multicolumn{1}{c}{\multirow{2}{*}{Constants}} & \multicolumn{1}{c|}{\multirow{2}{*}{$|\states_{\mc}|$}} & \multicolumn{1}{c}{\multirow{2}{*}{$|V_{\graph}|$}} & \multicolumn{1}{c}{\multirow{2}{*}{$\mathit{SCC}_{\graph}$}} & \multicolumn{1}{c}{\multirow{2}{*}{$\mathit{SCC}_{\mathit{pos}}$}} & \multicolumn{1}{c|}{\multirow{2}{*}{$T_{\graph}$}} & \multirow{2}{*}{$|\mathit{Vars}_{Z3}|$} & \multirow{2}{*}{$|\mathit{Cons}_{Z3}|$} & \multicolumn{1}{c|}{\multirow{2}{*}{$T_{\mathit{Z3}}$}} & \multirow{2}{*}{$T_{\mathit{mc}}$} \\ 
		\multicolumn{1}{c}{Property} & & & & & & & & & \\ 
		\midrule
		\multirow{8}{*}{
		\begin{tabular}{l}
		Crowds (TotalRuns, CrowdSize)\\
		\\
		$\logicProb\geq 0.9[ (\logicFinally \mathit{observe0}>1 \lor \logicGlobally \mathit{observe1}>1)$ \\
		$\land (\logicFinally \mathit{observe2}>1 \lor \logicGlobally\mathit{observe3}>1) ] $
		\end{tabular}
		}
		& 2,50 & 20534 & 82149 & 1374 & 0 & 4 & NE & NE & NE & 5 \\ 
		& 2,60 & 28446 & 113797 & 1940& 0 & 6  & NE & NE & NE& 7\\ 
		& 2,70 & 39131 & 156537 & 2615& 0 & 10  & NE & NE & NE & 11 \\
		& 2,80 & 51516 & 206077 & 3390& 0 & 21  & NE & NE& NE & 21 \\ 
		& 2,90 & 65601 & 262417 & 4265& 0 & 26  & NE & NE & NE& 27 \\ 
		& 2,100 & 81386 & 325557 & 5240&0 & 108  & NE & NE & NE& 108  \\
		& 2,110 & 98871 & 395497 & 6315&0 & 109 & NE & NE& NE & 110 \\ 
		& 2,120 & 118056 & 472237 & 7490&0& 118  & NE & NE& NE & 118 \\
		\midrule
		\multirow{8}{*}{
		\begin{tabular}{l}
		BRP (N, MAX)\\
		\\
		$\logicProb\geq 0.9[\logicFinally(\mathit{s}=5)\land\ltlFG(\mathit{rrep}=2)]$
		\end{tabular}
		}
		& 512,10 & 149976 & 893661& 1055 & 0 & 32 & NE & NE & NE & 33\\
		& 512,20 & 283476 & 1692861 & 1075 & 0 & 114 & NE & NE & NE & 114\\
		& 512,30 & 416976 & 2492061 & 1095 & 0 & 241 & NE & NE & NE & 241\\
		& 512,40 & 550476 & 3291261 & 1115 & 0 & 476 & NE & NE & NE & 477\\
		& 512,50 & 683976 & 4090461 & 1135 & 0 & 622 & NE & NE & NE & 623\\
		& 512,60 & 817476 & 4889661 & 1155 & 0 & 962 & NE & NE & NE & 962\\
		& 512,70 & 950976 & 5688861 & 1175 & 0 & 1219 & NE & NE & NE & 1220\\
		& 512,80 & 1084476 & MO & --  &  --   & -- & -- & -- & -- & --\\
		\midrule
		\multirow{8}{*}{
		\begin{tabular}{l}
		RPSC (N, L) \\
		\\
		$\logicProb\geq0.9[\logicFinally( \neg \mathit{knowA}) \lor\logicGlobally (\mathit{knowB})]$
		\end{tabular}
		}
		& 5,15  & 300030 & 1200117 & 4 & 1& 68 & 300031 & 300036 & 32 & 106 \\ 
		& 5,20 & 402430 & 1609717 & 4  & 1& 122 & 402431 & 402436 & 42 & 169 \\
		& 5,25  & 504830 & 2019317 & 4 & 1& 178 & 504831 & 504836  & 49 & 230 \\
		& 5,30 & 607230 & 2428917 & 4  & 1& 278 & 607231 & 607236 & 75& 373  \\ 
		& 5,35  & 709630 & 2838517 & 4 & 1& 384 & 709631 & 709636 & 86 & 488 \\ 
		& 5,40 & 812030 & 3248117 & 4  & 1& 490 & 812031 & 812036 & 95 & 600 \\ 
		& 5,45 & 914430 & 3657717 & 4  & 1& 606 & 914431 & 914436 & 104 & 719 \\
		& 5,50 & 1016830 & 4067317 & 4 & 1& 726 & 1016831 & 1016836 & 108 & 839 \\
		\bottomrule
		\end{tabular}
	}
\end{table}

As we can see from Tables~\ref{tab:paramtable} and~\ref{tab:paramtable2} relative to parametric Markov Chains, the implementation of the analysis method presented in this paper is able to check models in the order of millions of states. 
The model checking time is mainly depending on the behavior of Z3 and on the number and size of the SCCs of the product, since each SCC has to be classified as positive or negative;
regarding Z3, we can see that the time it requires for solving the provided system is loosely related to the size of the system:
the Crowds cases in Table~\ref{tab:paramtable} take hundreds of seconds for a system of size less than one thousand while the RPSC cases in Table~\ref{tab:paramtable2} are completed in less than one hundred seconds even if the system size is more than one million.
In Table~\ref{tab:imctable}, we list some experimental results for the Crowds protocol modelled as an interval Markov chain. 
As for the previous cases, it is the solution of the equation system to limit the size of the models we can analyse, so a more performing solver would improve considerably the applicability of our approach, in particular when we can not exclude that the formula is satisfiable, as happens when there are no positive SCCs.
\begin{table}[t]
  \centering
  \caption{Experimental Results for Crowds Interval Markov chain model}
  \label{tab:imctable}
  \resizebox{1\textwidth}{!}{%
        \scriptsize
        \setlength{\tabcolsep}{3pt}
		\begin{tabular}{ldr|rrrr|rrr|r}
		\toprule
		\multicolumn{1}{c}{Model (Constants)} & \multicolumn{1}{c}{\multirow{2}{*}{Constants}} & \multicolumn{1}{c|}{\multirow{2}{*}{$|\states_{\mc}|$}} & \multicolumn{1}{c}{\multirow{2}{*}{$|V_{\graph}|$}} & \multicolumn{1}{c}{\multirow{2}{*}{$\mathit{SCC}_{\graph}$}} & \multicolumn{1}{c}{\multirow{2}{*}{$\mathit{SCC}_{\mathit{pos}}$}} & \multicolumn{1}{c|}{\multirow{2}{*}{$T_{\graph}$}} & \multirow{2}{*}{$|\mathit{Vars}_{Z3}|$} & \multirow{2}{*}{$|\mathit{Cons}_{Z3}|$} & \multicolumn{1}{c|}{\multirow{2}{*}{$T_{\mathit{Z3}}$}} & \multirow{2}{*}{$T_{\mathit{mc}}$} \\ 
		\multicolumn{1}{c}{Property} & & & & & & & & & \\ 
		\midrule
		\multirow{6}{*}{
		\begin{tabular}{l}
		Crowds (TotalRuns, CrowdSize)\\
		\\
		$\logicProb\geq 0.9[ \ltlGF(\mathit{newInstance} \land $\\
		$\mathit{runCount}=0 \land \mathit{observe0} \ge 1)] $
		\end{tabular}
		}
		& 2,6 & 423 & 424 & 36     & 1&$<$1  & 240 & 473 & $<$1 & $<$1\\ 
		& 2,8 & 698 & 699 & 55     & 1&$<$1  & 380 & 760 & $<$1 & $<$1\\ 
		& 2,10 & 1041 & 1042 & 78  & 1&$<$1  & 552 & 1115 & 1 & 1\\ 
		& 2,12 & 1452 & 1453 & 105 & 1&$<$1  & 756 & 1538 & 15& 15 \\ 
		& 2,14 & 1931 & 1932 & 136 & 1&$<$1  & 992 & 2029 & 15& 15\\ 
		& 2,16 & 3093 & 3094 & 210 & 1&$<$1  & 1260 & 2588 & TO & TO \\ 

		\bottomrule
		\end{tabular}
  }
\end{table}

\section{Conclusion}
\label{sec:conclusion}
In this paper we have surveyed the parameter synthesis of PLTL formulas with respect to parametric Markov chains.
The algorithm first transforms the LTL specification to an almost fully partitioned automaton and then builds the product graph of the model under consideration
and this automaton.
Afterwards, we reduce the model checking problem to solving
an (nonlinear) equation system, which allows us employ an SMT solver to obtain feasible parameter values. 
We have conducted experiments to demonstrate that our techniques indeed work for models of realistic size.
To the best of our knowledge, our method is the first approach for the PLTL synthesis problem for parametric Markov chains which is single exponential in the size of the property.


\bibliographystyle{abbrv}

\begin{thebibliography}{10}

\bibitem{0020348}
C.~Baier and J.-P. Katoen.
\newblock {\em Principles of Model Checking}.
\newblock MIT Press, 2008.

\bibitem{BaierKKKMW16}
C.~Baier, S.~Kiefer, J.~Klein, S.~Kl\"uppelholz, D.~M\"uller, and J.~Worrell.
\newblock {Markov} chains and unambiguous {B\"uchi} automata.
\newblock In {\em CAV}, 2016.

\bibitem{BenediktLW13}
M.~Benedikt, R.~Lenhardt, and J.~Worrell.
\newblock {LTL} model checking of interval {Markov} chains.
\newblock In {\em TACAS}, volume 7795 of {\em LNCS}, pages 32--46, 2013.

\bibitem{BenediktLW14arxiv}
M.~Benedikt, R.~Lenhardt, and J.~Worrell.
\newblock Model checking {Markov} chains against unambiguous {B\"uchi}
  automata.
\newblock CoRR, available at~\url{http://arxiv.org/abs/1405.4560}, 2014.

\bibitem{BiancoA95}
A.~Bianco and L.~de~Alfaro.
\newblock Model checking of probabilistic and nondeterministic systems.
\newblock In {\em FSTTCS}, volume 1026 of {\em LNCS}, pages 499--513, 1995.

\bibitem{CartonM03}
O.~Carton and M.~Michel.
\newblock Unambiguous {B}{\"u}chi automata.
\newblock {\em TCS}, 297(1-3):37--81, 2003.

\bibitem{ChatterjeeGK13}
K.~Chatterjee, A.~Gaiser, and J.~Kret\'{\i}nsk{\'y}.
\newblock Automata with generalized {Rabin} pairs for probabilistic model
  checking and {LTL} synthesis.
\newblock In {\em CAV}, volume 8044 of {\em LNCS}, pages 559--575, 2013.

\bibitem{Chatterjee08}
K.~Chatterjee, K.~Sen, and T.~A. Henzinger.
\newblock Model-checking $\omega$-regular properties of interval {Markov}
  chains, 2008.

\bibitem{DBLP:conf/qest/CiesinskiB06}
F.~Ciesinski and C.~Baier.
\newblock {LiQuor}: A tool for qualitative and quantitative linear time
  analysis of reactive systems.
\newblock In {\em QEST}, pages 131--132, 2006.

\bibitem{Clarke08}
E.~M. Clarke.
\newblock The birth of model checking.
\newblock In {\em 25 Years of Model Checking}, volume 5000 of {\em LNCS}, pages
  1--26, 2008.

\bibitem{CGH94}
E.~M. Clarke, O.~Grumberg, and K.~Hamaguchi.
\newblock Another look at {LTL} model checking.
\newblock In {\em CAV}, volume 818 of {\em LNCS}, pages 415--427, 1994.

\bibitem{0007403}
E.~M. Clarke, O.~Grumberg, and D.~Peled.
\newblock {\em Model Checking}.
\newblock MIT Press, 2001.

\bibitem{CY95}
C.~Courcoubetis and M.~Yannakakis.
\newblock The complexity of probabilistic verification.
\newblock {\em J. of the ACM}, 42(4):857--907, 1995.

\bibitem{CSS03}
J.-M. Couvreur, N.~Saheb, and G.~Sutre.
\newblock An optimal automata approach to {LTL} model checking of probabilistic
  systems.
\newblock In {\em LPAR}, volume 2850 of {\em LNCS}, pages 361--375, 2003.

\bibitem{DArgenioJJL01}
P.~D'Argenio, B.~Jeannet, H.~Jensen, and K.~Larsen.
\newblock Reachability analysis of probabilistic systems by successive
  refinements.
\newblock In {\em PAPM/PROBMIV}, volume 2165 of {\em LNCS}, pages 39--56, 2001.

\bibitem{DeMoura08}
L.~De~Moura and N.~Bj{\o}rner.
\newblock Z3: An efficient smt solver.
\newblock In {\em TACAS}, TACAS'08/ETAPS'08, pages 337--340, Berlin,
  Heidelberg, 2008. Springer-Verlag.

\bibitem{Dehnert15}
C.~Dehnert, S.~Junges, N.~Jansen, F.~Corzilius, M.~Volk, H.~Bruintjes, J.-P.
  Katoen, and E.~{\'A}brah{\'a}m.
\newblock {PROPhESY: A PRObabilistic ParamEter SYnthesis Tool}.
\newblock In {\em CAV}, pages 214--231. Springer International Publishing,
  2015.

\bibitem{Esparza14}
J.~Esparza and J.~Kret\'{\i}nsk{\'y}.
\newblock {From LTL to Deterministic Automata: A Safraless Compositional
  Approach}.
\newblock In {\em CAV}, volume 8559 of {\em LNCS}, pages 192--208. Springer
  International Publishing, 2014.

\bibitem{Even85}
S.~Even, O.~Goldreich, and A.~Lempel.
\newblock A randomized protocol for signing contracts.
\newblock {\em Commun. ACM}, 28(6):637--647, June 1985.

\bibitem{HahnHZ11}
E.~M. Hahn, H.~Hermanns, and L.~Zhang.
\newblock Probabilistic reachability for parametric {Markov} models.
\newblock {\em STTT}, 13(1):3--19, 2011.

\bibitem{HahnLSTZ14}
E.~M. Hahn, Y.~Li, S.~Schewe, A.~Turrini, and L.~Zhang.
\newblock {IscasMC}: A web-based probabilistic model checker.
\newblock In {\em FM}, volume 8442 of {\em LNCS}, pages 312--317, 2014.

\bibitem{HaJo94}
H.~Hansson and B.~Jonsson.
\newblock A logic for reasoning about time and reliability.
\newblock {\em FAC}, 6(5):512--535, 1994.

\bibitem{HelminkSV94}
L.~Helmink, M.~Sellink, and F.~Vaandrager.
\newblock Proof-checking a data link protocol.
\newblock In {\em TYPES}, volume 806 of {\em LNCS}, pages 127--165, 1994.

\bibitem{Katoen12}
J.-P. Katoen, D.~Klink, M.~Leucker, and V.~Wolf.
\newblock {Three-valued abstraction for probabilistic systems}.
\newblock {\em The Journal of Logic and Algebraic Programming}, 81(4):356 --
  389, 2012.

\bibitem{KatoenZHHJ11}
J.-P. Katoen, I.~S. Zapreev, E.~M. Hahn, H.~Hermanns, and D.~N. Jansen.
\newblock The ins and outs of the probabilistic model checker {MRMC}.
\newblock {\em Performance Evaluation}, 68(2):90--104, 2011.

\bibitem{KemenySK66}
J.~G. Kemeny, J.~L. Snell, and A.~W. Knapp.
\newblock {\em Denumerable {M}arkov Chains}.
\newblock D. Van Nostrand Company, 1966.

\bibitem{Kini15}
D.~Kini and M.~Viswanathan.
\newblock {Limit Deterministic and Probabilistic Automata} for
  {LTL}{$\setminus$}{GU}.
\newblock In {\em TACAS}, volume 9035 of {\em LNCS}, pages 628--642, 2015.

\bibitem{KleinB07}
J.~Klein and C.~Baier.
\newblock On-the-fly stuttering in the construction of deterministic
  $\omega$-automata.
\newblock In {\em CIAA}, volume 4783 of {\em LNCS}, pages 51--61, 2007.

\bibitem{KwiatkowskaNP11}
M.~Kwiatkowska, G.~Norman, and D.~Parker.
\newblock {PRISM} 4.0: Verification of probabilistic real-time systems.
\newblock In {\em CAV}, volume 6806 of {\em LNCS}, pages 585--591, 2011.

\bibitem{LiuW09}
W.~Liu and J.~Wang.
\newblock A tighter analysis of {P}iterman's {B}{\"u}chi determinization.
\newblock {\em Information Processing Letters}, 109(16):941--945, 2009.

\bibitem{Piterman/07/Parity}
N.~Piterman.
\newblock From nondeterministic {B}\"uchi and {S}treett automata to
  deterministic parity automata.
\newblock {\em LMCS}, 3(3:5), 2007.

\bibitem{RR98}
M.~Reiter and A.~Rubin.
\newblock Crowds: Anonymity for web transactions.
\newblock {\em ACM TISSEC}, 1(1):66--92, 1998.

\bibitem{Safra/88/Safra}
S.~Safra.
\newblock On the complexity of {$\omega$}-automata.
\newblock In {\em FOCS}, pages 319--327, 1988.

\bibitem{Schewe/09/determinise}
S.~Schewe.
\newblock Tighter bounds for the determinisation of {B\"u}chi automata.
\newblock In {\em FoSSaCS}, volume 5504 of {\em LNCS}, pages 167--181, 2009.

\bibitem{Vardi85}
M.~Y. Vardi.
\newblock Automatic verification of probabilistic concurrent finite-state
  programs.
\newblock In {\em FOCS}, pages 327--338, 1985.

\end{thebibliography}

\clearpage
\appendix
\section*{Appendix}
\section{Proof of Theorem~\ref{thm:automata_cons}}
Here we only give the proof of the second claim. we denote by $\first(\varpi)$ the \emph{first} element of $\varpi$, i.e, $\first(\varpi) = w_{0}$.
\begin{proof}
According to the definition of transition function $\tranFunct_{\phi}$, we just need to prove $(V, \first(\pi))\satisfies \psi$ if and only if $\pi\models\psi$ for every $\pi \in \lang(\aut^U_{\phi})$, where $V\in\tranFunct_{\phi}(U, \first(\pi))$. Following we show the claim by induction on the structure of formula $\psi$.
\begin{enumerate}
\item $\psi=p\in\AP$. By satisfaction relation $\satisfies$, $(V, \first(\pi))\satisfies \psi$
is equivalent to $p\in \first(\pi)$ and further $\pi\models p$.
\item $\psi=\neg \psi_{1}$ or $\psi=\psi_{1}\land\psi_{2}$. Trivial by direct application of
induction hypotheses.
\item $\psi=\logicNext\psi_{1}$. Assume $\pi\models\logicNext\psi_{1}$ and $\pi \in \lang(\aut^U_{\phi})$, we have $\pi(1)\in\lang(\aut^V_{\phi})$ and $\pi(1)\models\psi_{1}$. By induction hypothesis $\logicNext\psi_{1}\in V$ if and only if $\pi(1)\models\psi_{1}$ for every $\pi(1)\in\lang(\aut^V_{\phi})$. Moreover, $\logicNext\psi_{1}\in V$ is equivalent to $(V, \first(\pi))\satisfies \logicNext\psi_{1}$, which completes the proof.
\item $\psi=\psi_{1}\logicUntil\psi_{2}$. We divide it into two cases:
\begin{enumerate}
\item $\pi\models\psi_{2}$. Trivial by induction hypothesis.
\item $\pi\models\psi_{1}\land\logicNext(\psi_{1}\logicUntil\psi_{2})$. By using above induction hypotheses
for both $\psi_{1}$ and $\logicNext(\psi_{1}\logicUntil\psi_{2})$, we complete the proof.
\end{enumerate}
\end{enumerate}
\end{proof}
\section{Proof of Corollary~\ref{cor:mutual_inf}}
\begin{proof}
Suppose $\sigma_{C}$ and its projection $\sigma_{K}$ is depicted as follows.
 \begin{figure}
	\centering
	\begin{tikzpicture}[->, >=stealth',shorten >=1pt,auto]
	
	\node (sigmaK) at (0, 0) {$\sigma_{K}$};
	\node (s0) at ($(sigmaK)+(1.2, 0)$) {$s_{0}$};
	\node (s1) at ($(sigmaK)+(3, 0)$) {$s_{1}$};
	\node (s2) at ($(sigmaK)+(4.8, 0)$) {$s_{2}$};
	\node (s3) at ($(sigmaK)+(6.1, 0)$) {$\cdots$};
	\node (si) at ($(sigmaK)+(7.73, 0)$) {$s_i$};
	\node (sip) at ($(sigmaK)+(9, 0)$) {$\cdots$};	
	
    \draw (s0) edge [] node {} (s1);
    \draw (s1) edge [] node {} (s2);
    \draw (s2) edge [] node {} (s3);
    \draw (s3) edge [] node {} (si);
    \draw (si) edge [] node {} (sip);

    	\node (sigmaC) at (0, 0.5) {$\sigma_{C}$};
	\node (q0) at ($(sigmaC)+(1, 0)$) {$(q_{0}, s_{0})$};
	\node (q1) at ($(sigmaC)+(2.8, 0)$) {$(q_{1}, s_{1})$};
	\node (q2) at ($(sigmaC)+(4.6, 0)$) {$(q_{2}, s_{2})$};
	\node (q3) at ($(sigmaC)+(6.1, 0)$) {$\cdots$};
	\node (qi) at ($(sigmaC)+(7.6, 0)$) {$(q_i, s_i)$};
	\node (qip) at ($(sigmaC)+(9, 0)$) {$\cdots$};	
	
    \draw (q0) edge [] node {} (q1);
    \draw (q1) edge [] node {} (q2);
    \draw (q2) edge [] node {} (q3);
    \draw (q3) edge [] node {} (qi);
    \draw (qi) edge [] node {} (qip);
    	
	\end{tikzpicture}
	\label{fig:completeSCCAndItsMap}
\end{figure}
\begin{enumerate}
\item $P_{C}\implies P_{K}$. Take any finite path $\sigma'_k$ of $K$, we can always get
a finite path $\sigma'_c$ of $C$ such that $\sigma'_k=\proj(\sigma'_c)$ since $C$
is complete. By assumption, $\sigma_{C}$ visits $\sigma'_c$ infinitely often and
$\sigma_{K}=\proj(\sigma_{C})$, then $\sigma_{K}$ must visit $\sigma'_k$ infinitely often.
\item $P_{K}\implies P_{C}$. Take any finite path $\sigma'_c$ of $C$. By Lemma~\ref{lem:prop_completeness}, we can find some finite path $\rho_k$ in $K$ (not necessarily that $\rho_k=\proj(\sigma'_c)$ holds)
such that with all $\rho_c$ of $C$ and $\rho_k = \proj(\rho_c)$, we have $\sigma'_c \trianglelefteq \rho_c$. According
to the assumption,
$\sigma_{K}$ visits $\rho_k$ infinitely often and $\sigma_{K}=\proj(\sigma_{C})$. It must be
the case that $\sigma_{C}$ visits some above finite path $\rho_c$ infinitely often and $\rho_c$
exists since $C$ is complete.
\end{enumerate}
\end{proof}
\section{Proof of Lemma~\ref{lem:comp_check}}
\begin{proof}
We show it by contradiction in the following.
\begin{enumerate}
\item $\ref{completeArcs})\Rightarrow\ref{firstSCCs})$. Suppose $C$ is complete and there exists an SCC $C'$ such that $\mathscr{H}(C') = \mathscr{H}(C)$ and there exists an arc from a state $(q_{0}, s)$ of $C'$ to a state $(q_{1}, t)$ of $C$. Since $\graph$ is reverse deterministic, the number of
$\labelFunc(s)$-predecessors of $(q_{1}, t)$ is at most one, thus the arc $(s,t)$ of $\mathscr{H}(C)$ does
not have a corresponding arc in $C$. Therefore $\ref{completeArcs})$ does not hold.
\item $\ref{firstSCCs})\Rightarrow\ref{completeArcs})$. Suppose $C$ is an SCC and there exists some arc $(s, t)$ in $\mathscr{H}(C)$ which
does not have a corresponding arc in $C$. We first find a state, say $(q, t)$ in $C$, and get its
immediate $\labelFunc(s)$-predecessor $(p, s)$, which is obviously not in $C$ since $\graph$ is reverse-deterministic. We assume that $C'$ is the SCC contains state $(p, s)$ with $\mathscr{H}(C') = \mathscr{H}(C)$, which leads to $C' \preceq C$. Thus we complete the proof.
\end{enumerate}
\end{proof}
\end{document}